\documentclass[aps,prb,reprint,superscriptaddress]{revtex4-2}
\usepackage{amsmath,amssymb,amsfonts}
\usepackage{graphicx}
\usepackage{bm}
\usepackage{hyperref}
\usepackage{booktabs}
\usepackage{subcaption}

\newcommand{\Jone}{J_1}
\newcommand{\Jz}{J_{\perp}}
\newcommand{\ta}{t_a}
\newcommand{\Geff}{\Gamma_{\mathrm{eff}}}
\newcommand{\mumon}{\mu_{\rm mon}}
\newcommand{\kstar}{\mathbf{k}^{*}}
\newcommand{\rhom}{\rho_m}

\begin{document}


\title{Quantum simulation of interlayer charge ordering in Kagome
frustrated-magnet}

\author{Kumar Ghosh}
\email{jb.ghosh@outlook.com}
\affiliation{E.ON Digital Technology, Laatzener Str.\ 1,
             30539 Hannover, Germany}
\author{Sabre Kais}
\email{skais@ncsu.edu}
\affiliation{Department of Electrical and Computer Engineering, North Carolina State University, Raleigh, NC 27606}
\affiliation{ Department of Chemistry, and Department of Physics, North Carolina State University, Raleigh, North Carolina 27695}


\begin{abstract}
The interplay between interlayer coupling and quantum fluctuations governs charge ordering and defect dynamics in kagome systems, yet these parameters are intrinsically entangled in existing kagome metals and artificial spin-ice platforms, preventing their independent control. Here we realize a bilayer kagome frustrated-magnet simulator comprising 1,536 connected spins on a D-Wave Advantage2 quantum annealer, in which the effective quantum drive, $\Gamma_{\rm eff}$,  and interlayer exchange, $J_{\perp}$, are independently programmable. We observe an interlayer-driven transition from ferroelectric to antiferroelectric Ice-II charge order at a critical coupling $(J_\perp/J_1)^*\approx0.04$,  a phenomenon absent in single-layer geometries and robust across a wide range of annealing conditions. Classical Monte Carlo calculations establish that the transition persists in the classical limit, allowing the experimentally observed critical coupling to quantify the quantum renormalization induced by fluctuations. Applying the resulting phase diagram to kagome charge-density-wave materials placesKV$_3$Sb$_5$ and RbV$_3$Sb$_5$ deep within the ordered antiferroelectric regime, while locating CsV$_3$Sb$_5$ near the phase boundary, providing a natural explanation for its metastable $2\times2\times4$ stacking order. We further show that restricting charge-correlation measurements to ice-rule configurations resolves a systematic underestimation of ordering in conventional analyses, enabling direct reinterpretation of resonant X-ray, XMCD, STM and anomalous Hall experiments. Finally, we demonstrate that the same charge-sector reorganization framework explains near-degenerate plateau states in the metallic kagome spin-ice HoAgGe and yields experimentally testable predictions for nanomagnetic, kagome-metal and van der Waals frustrated systems. These results establish programmable quantum annealers as scalable simulators of emergent charge order and monopole physics in frustrated quantum matter.
\end{abstract}

\keywords{quantum annealing, artificial spin ice, kagome lattice,
magnetic monopoles, bilayer frustrated magnetism, Ice-II, D-Wave}

\maketitle

\section{Introduction}
\label{sec:intro}

Layered kagome metals display a striking diversity of charge-density
wave (CDW) ground states whose three-dimensional character is governed
by interlayer stacking. In the AV$_3$Sb$_5$ family ($A=$ K, Rb, Cs),
K and Rb variants adopt a uniform $2\times2\times2$ trihexagonal
phase, whereas CsV$_3$Sb$_5$ develops a metastable $2\times2\times4$
mixed-layer pattern whose stability is not captured by any single-layer
model~\cite{Kautzsch2023,Wilson2024}. Synchrotron X-ray refinements
resolve that KV$_3$Sb$_5$ and RbV$_3$Sb$_5$ share a staggered
trihexagonal (TrH) ground state in space group $Fmmm$ with
$c=8.845$\,\AA{} and $9.107$\,\AA{} respectively, while the larger
Cs interlayer ($c=9.362$\,\AA{}) drives a metastable $2\times2\times4$
superstructure with mixed TrH and Star-of-David (SoD) layers and
a staged onset: a $2\times2\times2$ phase appears first near
$T_{\mathrm{CDW}}\approx94$\,K, followed by a $2\times2\times4$ state
below $\approx85$\,K with scattering weight exchanged between the two
order parameters~\cite{Kautzsch2023}. Why continuous alkali-cation
substitution, which monotonically tunes interlayer spacing by
$\approx0.26$\,\AA{} per step, produces qualitatively distinct CDW
ground states rather than a smooth evolution remains the central
unresolved question in this materials class~\cite{Wilson2024,Wang2023}.
In the bilayer kagome metal ScV$_6$Sn$_6$, two CDW wavevectors with
distinct out-of-plane components, $q_3=(\tfrac{1}{3},\tfrac{1}{3},\tfrac{1}{3})$
and $q_2=(\tfrac{1}{3},\tfrac{1}{3},\tfrac{1}{2})$,
compete~\cite{Arachchige2022}, with no resolved mechanism selecting
between them. Inelastic X-ray scattering has now established that the
$q_2$ instability is the dominant phonon-softening channel above
$T_{\rm CDW}$, while the $q_3$ state that condenses at $92$\,K
is selected by a wavevector-dependent electron-phonon
coupling~\cite{Cao2023}. The $q_3$ ground state is accompanied by a
$+0.04\%$ c-axis expansion at the $92$\,K transition, signalling a
reduction in effective interlayer coupling precisely when CDW order
sets in~\cite{Arachchige2022}. In CT-MoTe$_2$, mirror-twin-boundary
superatoms form kagome-like flat and Dirac bands whose interlayer
charge correlations are uncharacterised~\cite{LeiMoTe2023}. In
Fe$_3$Sn$_2$, ABA-stacked bilayer kagome produces a Berry-curvature
Hall response absent in the monolayer~\cite{Ye2018}. Across this
entire materials class, the same design limitation recurs: the
interlayer coupling governing charge-layer stacking and the quantum
drive controlling defect density are structurally entangled and cannot
be varied independently.

The same limitation extends to metallic frustrated magnets. HoAgGe
crystallises in space group $P\bar{6}2m$ with Ho$^{3+}$ ions forming a
distorted kagome lattice in the $ab$-plane, where two distinct triangle
types are rotated by opposite angles of $\approx15.58^{\circ}$ around
the $c$-axis~\cite{Zhao2020}. This is the first crystalline (non-artificial)
system confirmed to realise the kagome spin-ice state, with the Ho
moments governed by dominant nearest-neighbour ferromagnetic exchange
and RKKY-type further-neighbour couplings extending to the 4th nearest
neighbour, which are found to be stronger than the long-range dipolar
interaction~\cite{Zhao2020}. The fully ordered $\sqrt{3}\times\sqrt{3}$
ground state below $\approx4$\,K carries a residual entropy of
$\approx0.501\,k_B$ per spin~\cite{Zhao2020}, identical to the
classical kagome spin-ice entropy realised in our simulator at
$\Gamma=J_\perp=0$. Under in-plane field, the system passes through
1/3, 2/3, and 5/6 magnetization plateaus through metamagnetic
transitions that preserve the kagome ice rule. Zhao~et~al.\ subsequently
established through anomalous Hall effect (AHE) measurements that these
plateau phases host at least two near-degenerate ice-rule states,
S$_{1/3}$ and S$'_{1/3}$, with identical net magnetization and identical
band structures but different Berry curvatures and consequently different
anomalous Hall conductivities~\cite{Zhao2024}. This Berry-curvature
contrast arises from a time-reversal-like symmetry operation that
reverses the $15.58^{\circ}$ lattice distortion while leaving spin
degrees of freedom unchanged, thereby shifting the Berry connection
(position-operator matrix elements between Bloch states) without
altering eigenvalues~\cite{Zhao2024}. The physical consequence is that
the magnetization shows vanishing hysteresis across the plateaus while
the AHE exhibits pronounced finite hysteresis at 2\,K~\cite{Zhao2024}.
This constitutes a charge-sector near-degeneracy on the ice manifold
that is exactly the transport fingerprint of proximity to a staggered
charge-ordering boundary, yet no platform exists in which the interlayer
coupling selecting between these near-degenerate configurations and the
quantum drive controlling defect density can be varied independently so
as to map the full phase boundary. The bilayer kagome simulator
introduced here resolves this impasse for both the CDW and the metallic
spin-ice contexts simultaneously.

The same constraint applies to artificial kagome platforms. In
Permalloy nanomagnet arrays, PEEM and MFM imaging confirm
exponentially decaying Dirac-string statistics~\cite{Ladak2010,Mengotti2011},
but the monopole chemical potential is fixed by vertex geometry. In
the three-dimensional Ni$_{81}$Fe$_{19}$ diamond-bond nanowire lattice
of May~et~al.~\cite{May2019} ($r=80$\,nm, arc length $\ell_{\rm eff}
\approx330$\,nm), finite-element simulations show that type-1 and
type-2 vertex energies agree to within $10\%$, placing the system
close to the degenerate Coulomb ice point, but the interlayer
separation $d_z$ is fixed by the two-photon lithography template and
cannot be varied continuously. In bulk pyrochlore spin ice, neutron
scattering establishes the magnetic Coulomb
phase~\cite{Castelnovo2008,Morris2009,Fennell2009}, but thermal and
quantum fluctuations are coupled to the same microscopic exchange.
Bridged kagome~\cite{Hofhuis2022} and direct-kagome~\cite{Yue2024}
artificial spin-ice systems each offer only a single tunable coupling;
in particular, the theoretical prediction of a charge-ordered kagome Ice-II
phase via a two-stage ordering sequence~\cite{Chern2011} has resisted
experimental confirmation in these single-layer geometries because the
relevant charge ordering requires a bilayer probe.

Here we resolve this impasse with a bilayer kagome frustrated-magnet
simulator implemented on the D-Wave Advantage2 Zephyr processor,
which maps onto two coupled kagome planes via \textsc{minorminer}
embedding with zero chain-break fractions across all conditions.
The frustrated bilayer kagome transverse-field Ising model (TFIM)
carries a sign problem for $\Gamma>0$~\cite{Troyer2005} and the
bilayer geometry imposes bond-dimension growth scaling as
$\exp(\alpha L)$, making exact classical simulation intractable at
$N=1{,}536$ spins. The quantum processing unit provides $\Geff$ and
$\Jz$ as genuinely independent programmable material parameters over a
two-dimensional phase space with no analogue in existing solid-state
or artificial-nanomagnet realisations.
Fig.~\ref{fig:materials_map} places the four target materials systems
on our two-dimensional phase diagram, establishing the quantitative
framework within which their CDW and spin-ice phenomenology is
interpreted below.

\section*{Results}

\begin{figure*}[t]
  \centering
  \includegraphics[width=0.95\textwidth]{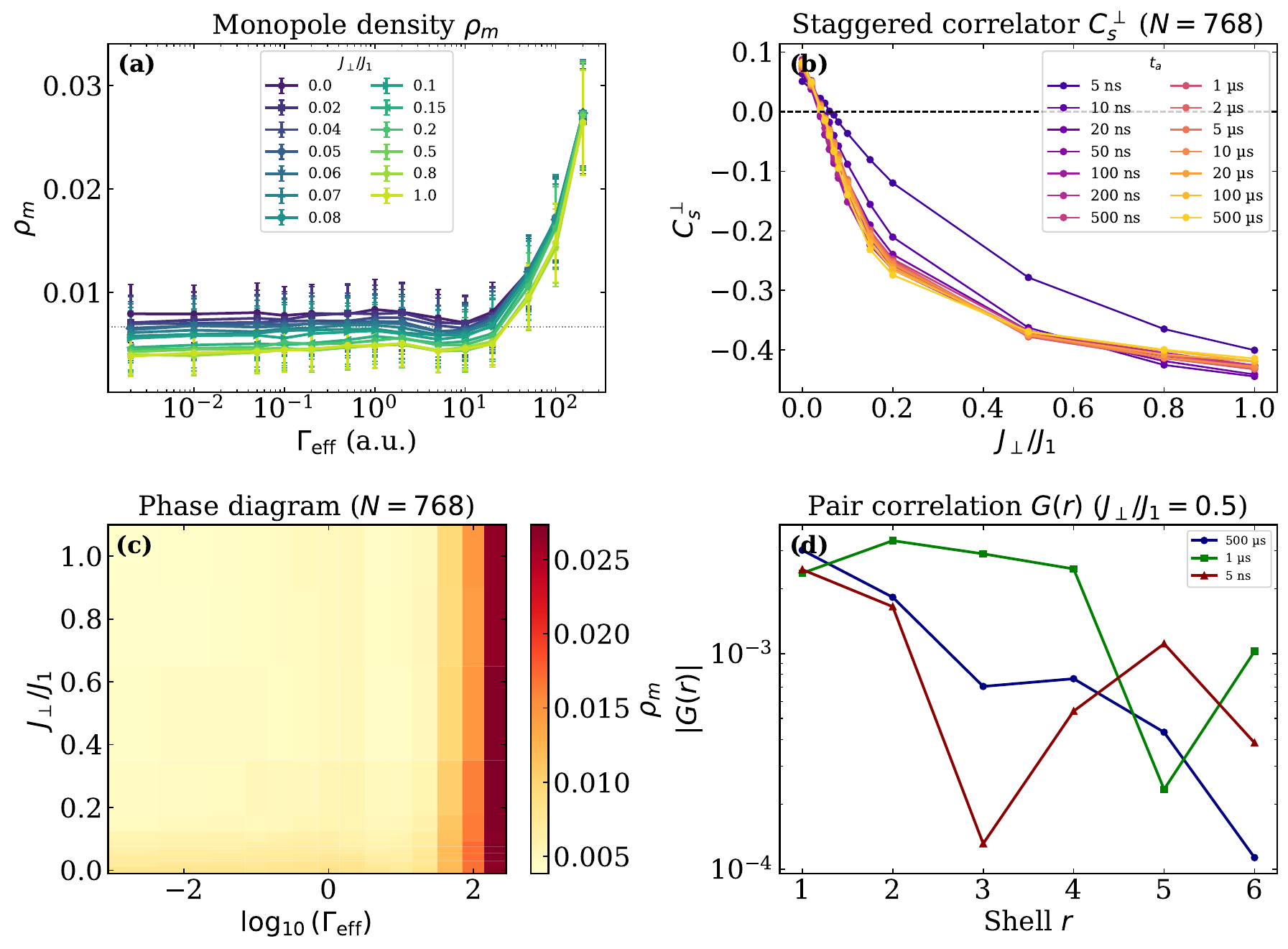}
  \caption{Independently programmable charge order and monopole density
  at $N=768$/layer.
  \textbf{(a)} Monopole density $\rhom$ versus $\Geff$ for all
  $\Jz/\Jone$ values; collapse across all thirteen interlayer couplings
  confirms $\Geff$ as the sole driver of monopole activation.
  \textbf{(b)} $C_s^\perp$ [Eq.~\eqref{eq:csperp}] versus $\Jz/\Jone$
  for all annealing times; collapse across all $\Geff$ confirms $\Jz$
  as the sole driver of staggered charge ordering.
  \textbf{(c)} Phase diagram of $\rhom$ in
  $(\log_{10}\Geff,\,\Jz/\Jone)$ space: vertical $\rhom$ stripes and a
  horizontal charge-ordering boundary confirm the two-parameter
  factorisation.
  \textbf{(d)} Monopole pair correlation $G(r)$ at three representative
  annealing times ($\Jz/\Jone=0.5$); exponential decay confirms the
  confined phase.}
  \label{fig:composite_4panel}
\end{figure*}

\paragraph{Independently programmable charge order and monopole density.}
The central platform result is a factorisation: monopole density
$\rhom$ responds exclusively to $\Geff$ and the interlayer staggered
charge correlator $C_s^\perp$ responds exclusively to $\Jz$, across
the full two-dimensional parameter space.
Fig.~\ref{fig:composite_4panel} establishes this independence.
$\rhom$ rises monotonically by a factor of seven as $\Geff$ increases,
with collapse across all thirteen interlayer couplings (panel~a),
confirming $\Geff$ as the sole driver of monopole activation.
Conversely, $C_s^\perp$ reverses sign from weakly ferroelectric
($\approx+0.05$ to $+0.09$ at $\Jz=0$) to strongly antiferroelectric
($\approx-0.40$ to $-0.44$ at $\Jz/\Jone=1$) with collapse across
all annealing times (panel~b), confirming $\Jz$ as the sole driver of
staggered charge ordering. The two-dimensional phase diagram (panel~c)
makes the factorisation explicit: $\rhom$ stripes run vertically and
the charge-ordering boundary runs horizontally, confirming that tuning
one parameter leaves the other unchanged. This independence, absent in
any previous frustrated-magnet platform, is the prerequisite for
quantitative materials design of the monopole phase diagram.
Panel~(d) shows the monopole pair correlation $G(r)$ decaying
exponentially; the confinement length $\xi$ increases with $\Geff$,
signalling progressive Dirac-string softening without the divergence
expected at deconfinement. Three independent confinement diagnostics,
exponential $G(r)$ decay, universally negative Bayesian model
selection ($\Delta\mathrm{BIC}\in[-4.05\times10^3,-6.3\times10^2]$),
and $\rho_{\max}<1$, confirm the confined phase throughout
(Supplementary Information).

\paragraph{An interlayer-exchange-driven charge-ordering instability
absent in single-layer kagome geometries.}
Independent tunability of $\Jz$ directly reveals a charge-ordering
instability with no single-layer analogue, one that maps quantitatively
onto the stacking competition in vanadium-based kagome metals. The
interlayer staggered correlator $C_s^\perp$ reverses sign in the
window $(\Jz/\Jone)^*\approx0.04$--$0.05$, signalling a
ferroelectric-to-antiferroelectric Ice-II transition. This is the
interlayer generalisation of the Moller-Moessner Ice-II
crystal~\cite{Moller2009}: when $\Jz>0$ couples two kagome planes,
their NaCl sublattices develop opposite staggered charge order
stabilised by interlayer exchange, a configuration invisible to all
single-layer probes because $C_s^\perp$ [Eq.~\eqref{eq:csperp}]
couples charges across two distinct planes. The single-layer kagome Ice-II
transition, first predicted by Chern, Mellado, and Tchernyshyov to
belong to the 2D Ising universality class~\cite{Chern2011}, is the
single-layer limit of the transition we observe here; the bilayer
geometry adds a strictly new channel, the interlayer staggered order,
that is absent in all existing artificial kagome realisations~\cite{Hofhuis2022,Yue2024}.

\begin{figure*}[htbp]
\centering
\includegraphics[width=0.95\textwidth]{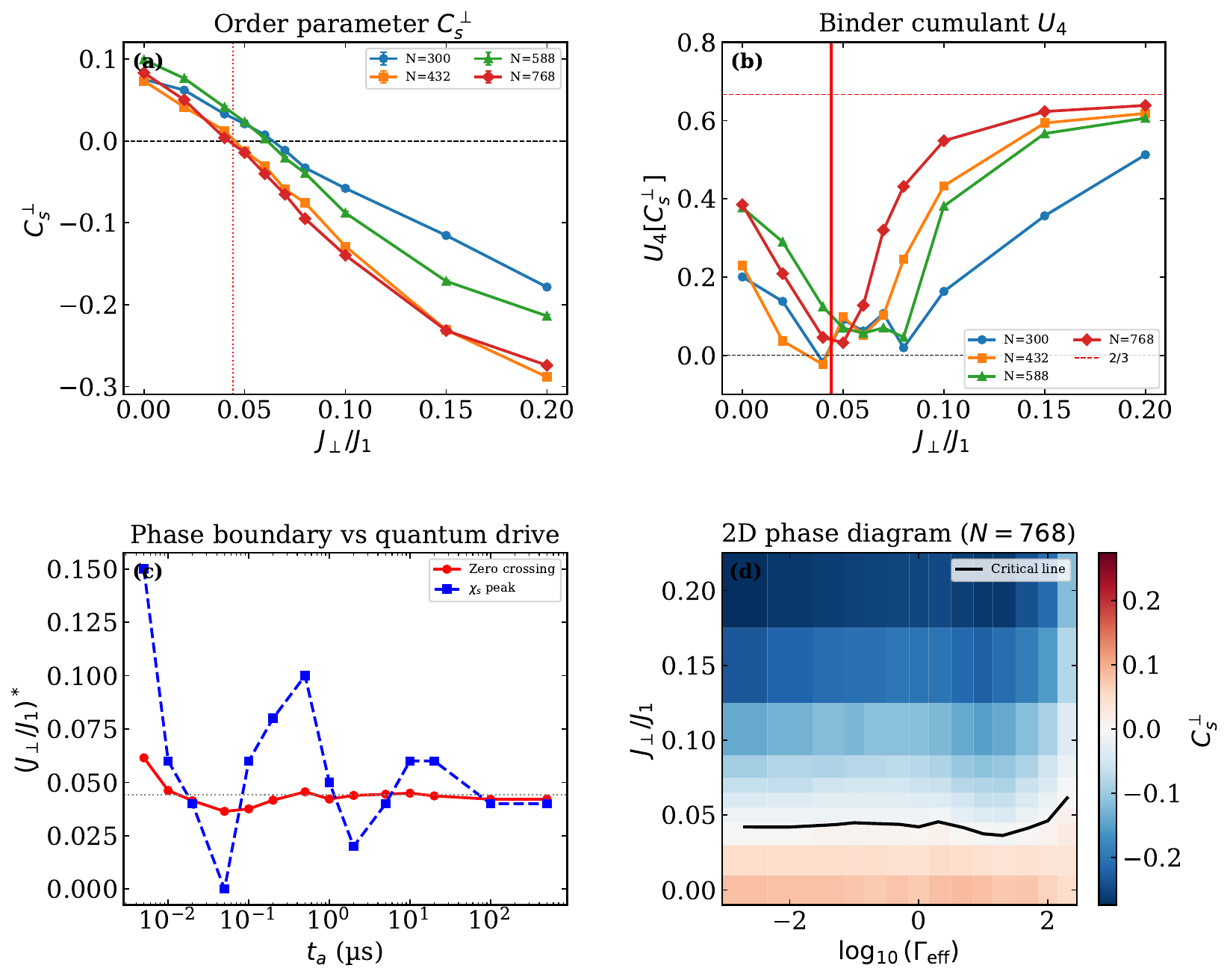}
\caption{Interlayer-exchange-driven ferroelectric-to-antiferroelectric
Ice-II charge-ordering instability.
\textbf{(a)} Slow-anneal order-parameter curves $C_s^\perp$
[Eq.~\eqref{eq:csperp}] for all four system sizes, showing the direct
sign reversal near $(\Jz/\Jone)^*\approx0.04$.
\textbf{(b)} Slow-anneal Binder cumulants $U_4[C_s^\perp]$ crossing
in the same window; the dashed line at $U_4=2/3$ marks the
ordered-phase limit.
\textbf{(c)} Phase boundary versus annealing time $\ta$ from two
independent estimators: the zero crossing of $C_s^\perp$ (filled
squares) and the peak of its susceptibility $\chi_s$ (open squares).
Both remain stable within $0.04$--$0.05$ over five decades; the
upward shift only at the non-adiabatic $5\,\mathrm{ns}$ point confirms
that the critical coupling is a ground-state property of the bilayer
exchange.
\textbf{(d)} Two-dimensional phase diagram in
$(\log_{10}\Geff,\,\Jz/\Jone)$ space at $N=768$; the horizontal
critical line confirms the charge-ordering boundary is independent of
quantum drive.}
\label{fig:transition_composite}
\end{figure*}

Fig.~\ref{fig:transition_composite} presents four independent
diagnostics. The zero crossing of $C_s^\perp$ across all four system
sizes at slow anneal ($\ta=500\,\mu$s) falls near
$(\Jz/\Jone)^*\approx0.04$ (panel~a), corroborated by the Binder
cumulant $U_4[C_s^\perp]$ crossing in the same window (panel~b). The
zero-crossing is stable within $0.04$--$0.05$ from $10\,\mathrm{ns}$
to $500\,\mu\mathrm{s}$ (panel~c): only the non-adiabatic
$5\,\mathrm{ns}$ point shifts upward, attributable to hardware
freeze-out rather than quantum renormalisation of the critical coupling.
The two-dimensional phase diagram (panel~d) confirms a strictly
horizontal critical line, establishing that the transition coupling is
independent of quantum drive and is therefore a ground-state property
of the bilayer exchange. A finite-size scaling scan is compatible with
a continuous transition ($\nu\gtrsim1$); the analogous single-layer
transition belongs to the 2D Ising universality class~\cite{Chern2011,Hofhuis2022},
and whether interlayer coupling shifts this class is a concrete
prediction testable on next-generation hardware with a factor of
$\approx2$--$3$ increase in linear dimension $L$.

Blind holdout validation confirms the robustness of the critical
coupling: all five withheld-point predictions remain sub-$1\sigma$,
with the most stringent near-critical test ($\Jz/\Jone=0.04$ withheld)
deviating by only $0.62\sigma$ (Supplementary Information),
inconsistent with first-order interlayer reordering. The transition is
a charge-sector reorganisation of the ice manifold rather than a
monopole-proliferation event: staggered composite monopoles (plaquettes
carrying opposite-sign defects in both layers simultaneously) remain
three orders of magnitude more dilute than standard monopoles at the
critical point ($\rho_{\mathrm{sc}}/\rhom\approx2.7\times10^{-3}$;
Supplementary Fig.~S8).
Classical parallel-tempering Monte Carlo simulations of the
$\Gamma=0$ limit confirm that the charge-ordering transition exists
classically at $J_\perp=0$ by $\mathbb{Z}_2$ symmetry; the QPU
critical coupling $(J_\perp/J_1)^*\approx0.04$ therefore constitutes
a direct measure of quantum renormalisation of the critical coupling
by the transverse field $\Geff$ (Supplementary Note, Fig.~S10).

\paragraph{A recalibrated charge-structure-factor protocol applicable
to kagome-metal, spin-ice, and anomalous Hall datasets.}
The independent separation of defect and non-defect plaquettes exposes
a systematic underestimation of charge-order amplitude in all published
kagome and spin-ice datasets where monopole or defect-site density is
non-negligible. The conventional all-plaquette structure factor
$S_Q(\kstar)$ sums over every plaquette regardless of ice-rule
compliance; defect plaquettes ($|Q_m|=\tfrac{3}{2}$) carry incoherent
charge that dilutes the Ice-II signal in proportion to the defect
fraction. The corrected estimator $S_Q^{\mathrm{stag}}(\kstar)$
[Supplementary Eq.~(S6)] restricts the sum to ice-rule plaquettes,
eliminating this dilution. This separation is structurally identical
to Yue~et~al.'s decomposition of spin and toroidal-moment magnetic
structure factors in direct-kagome artificial spin ice~\cite{Yue2024}.

\begin{figure*}[t]
  \centering
    \begin{minipage}[t]{0.49\textwidth}
        \centering
        \includegraphics[width=\textwidth]{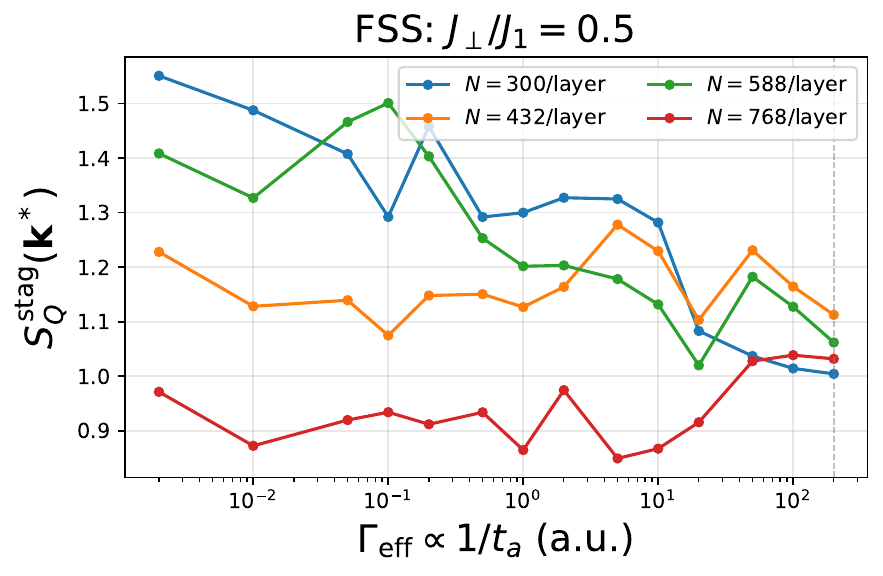}
    \end{minipage}
    \hfill
    \begin{minipage}[t]{0.49\textwidth}
        \centering
        \includegraphics[width=\textwidth]{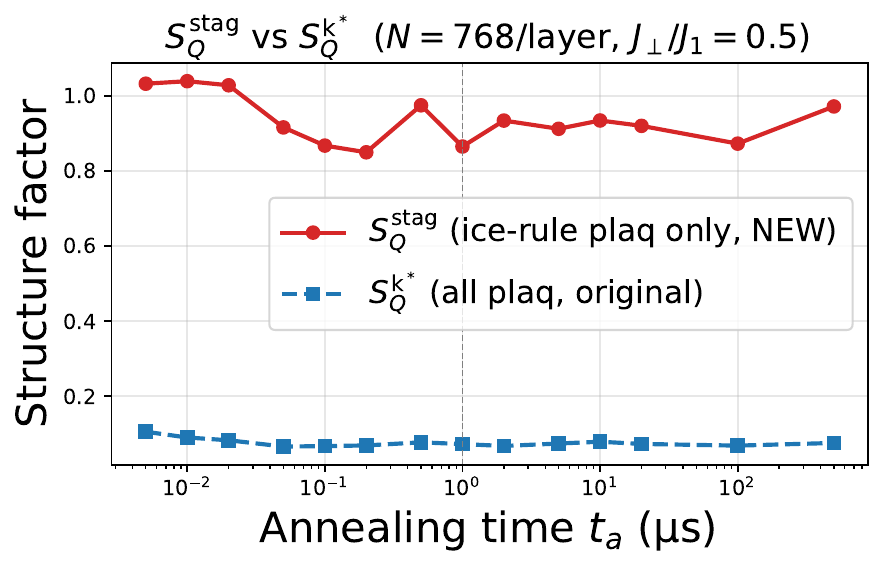}
    \end{minipage}
  \caption{Recalibrated charge-structure-factor protocol for
  frustrated-magnet, kagome-metal, and anomalous Hall datasets.
  \emph{Left:} $S_Q^{\mathrm{stag}}(\kstar)$ versus $\Geff$ for four
  system sizes ($N=300$, $432$, $588$, $768$/layer) at $\Jz/\Jone=0.5$;
  $N$-dependent growth at fast anneals identifies the
  quantum-to-classical crossover and confirms quantum-driven Ice-II
  order selection.
  \emph{Right:} $S_Q^{\mathrm{stag}}(\kstar)$ versus the conventional
  all-plaquette $S_Q(\kstar)$ for $N=768$, $\Jz/\Jone=0.5$, as a
  function of $\ta$. The order-of-magnitude correction ($9.8$--$13.9\times$,
  typical $\approx12$) demonstrates systematic underestimation by
  conventional probes whenever defect density is non-negligible,
  including in published anomalous Hall measurements of metallic kagome
  spin-ice systems~\cite{Zhao2024}.}
  \label{fig:sqcomp}
\end{figure*}

Fig.~\ref{fig:sqcomp} quantifies the correction at $N=768$,
$\Jz/\Jone=0.5$: $S_Q^{\mathrm{stag}}(\kstar)$ exceeds the
all-plaquette $S_Q(\kstar)$ by a factor of $9.8$--$13.9$ (typical
$\approx12$). This correction applies whenever $\rhom\gtrsim5\%$,
which holds throughout the fast-anneal regime and across a large
fraction of published spin-ice and kagome-metal
datasets~\cite{King2021,LopezBezanilla2023,May2019,Farhan2019}. Applying
$S_Q^{\mathrm{stag}}(\kstar)$ to published XMCD or MFM vertex maps,
including those of Refs.~\cite{May2019,Yue2024}, is predicted to
recover Ice-II charge-order amplitude $\approx12\times$ above
previously reported values from data already in hand. The same
ice-rule restriction applies directly to the anomalous Hall datasets of
HoAgGe~\cite{Zhao2024}: within each field-induced magnetization plateau
the monopole (defect-plaquette) density is non-negligible, so the
all-site Hall resistivity measurement conflates the coherent
Berry-curvature contribution from ice-rule plaquettes with incoherent
defect-site contributions. Restricting the charge-sector analysis to
ice-rule plaquettes using $S_Q^{\mathrm{stag}}(\kstar)$ predicts
recovery of the Berry-curvature contrast between the near-degenerate
states S$_{1/3}$ and S$'_{1/3}$ at sensitivity $\approx12\times$
above the conventional all-site AHE estimator, making the
time-reversal-like symmetry distinction between them measurable in
published Hall data without new crystal growth or device
fabrication~\cite{Zhao2024}. The
$N$-dependent growth of $S_Q^{\mathrm{stag}}$ at fast anneals
($\ta\lesssim1\,\mu$s), absent at slow anneals, is a finite-size
signature of quantum-driven Ice-II order selection extending the
single-layer mechanism of Ref.~\cite{LopezBezanilla2023} to the
bilayer geometry.

\paragraph{Falsifiable predictions for four materials platforms.}
The two-parameter phase diagram and the recalibrated estimator yield
concrete, falsifiable predictions for four materials platforms, each
testable against existing experimental data without new sample
preparation. Fig.~\ref{fig:materials_map} places all four systems
quantitatively on the $(J_\perp/J_1,\,\Geff)$ phase diagram.

\emph{AV$_3$Sb$_5$ kagome superconductors.}
In the AV$_3$Sb$_5$ family, the alkali cation sets the c-axis lattice
parameter and hence the effective interlayer V-V exchange, providing a
direct chemical realisation of the $\Jz/\Jone$ axis. Single-crystal
synchrotron refinements establish c-axis parameters of $8.845$\,\AA{}
(K), $9.107$\,\AA{} (Rb), and $9.362$\,\AA{} (Cs) at room
temperature~\cite{Kautzsch2023}, a monotone expansion of
$\approx0.26$\,\AA{} per alkali substitution that directly maps onto
decreasing effective $\Jz/\Jone$. Our phase diagram predicts that
KV$_3$Sb$_5$ and RbV$_3$Sb$_5$, with their smaller interlayer
spacings~\cite{Wilson2024}, sit above $(J_\perp/J_1)^*\approx0.04$
and carry a well-defined staggered antiferroelectric interlayer
charge-order pattern, while CsV$_3$Sb$_5$ sits near the boundary
where competing stacking orders become nearly degenerate.

This interpretation quantitatively explains the staged ordering
behaviour in CsV$_3$Sb$_5$ revealed by X-ray temperature clustering
(X-TEC) analysis~\cite{Kautzsch2023}: at $90$\,K a $2\times2\times2$
phase appears first, followed by the $2\times2\times4$ state below
$85$\,K with scattering weight exchanged between the two order
parameters. This staging---with the $2\times2\times2$ peaks initially
\emph{decreasing} upon onset of the $2\times2\times4$ peaks and then
partially recovering---is a critical-fluctuation signature of proximity
to $(J_\perp/J_1)^*$, not a property unique to the Cs compound. Below
the critical coupling, competing stacking orders are nearly degenerate,
the system cannot commit to a single phase upon cooling, and the
observed thermal history and quench-rate sensitivity of the ground
state follow naturally. Above the critical coupling, as in K and Rb
variants, the antiferroelectric stacking is unambiguously selected and
no staging is observed. The reviews by Wang~et~al.~\cite{Wang2023} and
Wilson and Ortiz~\cite{Wilson2024} identify the origin of this
qualitative jump in CDW stacking across the alkali series as an open
question; our result provides a mechanistic answer in terms of a
genuine phase boundary rather than a smooth crossover. The
V-V bond-length standard deviations in the CDW state
($\pm0.022$\,\AA{} for K, $\pm0.028$\,\AA{} for Rb, $\pm0.036$\,\AA{}
for Cs~\cite{Kautzsch2023}) grow monotonically, directly reflecting
the increasing proximity to the phase boundary. Intermediate-temperature
anomalies in CsV$_3$Sb$_5$ near $T_a\approx60$\,K and
$T_b\approx35$\,K reported across multiple probes~\cite{Wilson2024}
may therefore reflect critical fluctuations of the interlayer stacking
order parameter, rather than independent electronic instabilities.

A directly testable prediction follows: resonant X-ray scattering on
the K-Rb-Cs series, analysed with $S_Q^{\mathrm{stag}}(\kstar)$
restricted to non-defect V sites---the direct analogue of our
ice-rule-restricted estimator applied to vanadium kagome
plaquettes---should reveal a monotone decrease in staggered
charge-order amplitude from K to Cs, with an absolute correction
factor of $\approx12\times$ relative to the conventional all-site
estimator. This test requires no new crystal growth and is feasible
against the synchrotron datasets already published in
Ref.~\cite{Kautzsch2023}.

\emph{ScV$_6$Sn$_6$ bilayer kagome and the competing-wavevector problem.}
The competing CDW wavevectors
$q_2=(\tfrac{1}{3},\tfrac{1}{3},\tfrac{1}{2})$ and
$q_3=(\tfrac{1}{3},\tfrac{1}{3},\tfrac{1}{3})$ have distinct
out-of-plane components whose relative stability is governed by
interlayer coupling~\cite{Arachchige2022}. Inelastic X-ray scattering
by Cao~et~al.~\cite{Cao2023} establishes that $q_2$ develops as a
short-range dynamic instability above $T_{\rm CDW}$, while the $q_3$
state condenses as the long-range ground state at $92$\,K, with
the two orders in direct competition. Our bilayer phase diagram
provides a transparent geometric interpretation of this hierarchy:
the $q_3$ state, which has the longer c-axis repeat and weaker
interlayer modulation, corresponds to the ferroelectric ($C_s^\perp>0$)
side of our transition, while $q_2$, with its shorter out-of-plane
repeat and stronger interlayer modulation, corresponds to the
antiferroelectric ($C_s^\perp<0$) side. Crucially, the $+0.04\%$
c-axis expansion observed at the $92$\,K transition~\cite{Arachchige2022}
means that the effective interlayer coupling \emph{weakens} precisely
when the CDW sets in, self-consistently stabilising the $q_3$
ferroelectric state. ScV$_6$Sn$_6$ therefore sits close to but below
$(J_\perp/J_1)^*$ at ambient pressure: the short-range $q_2$
fluctuations observed above $T_{\rm CDW}$~\cite{Cao2023} correspond
to the system sampling the antiferroelectric side of the transition
before the c-axis expansion at $T_{\rm CDW}$ pushes it back below
the critical coupling.

Applied hydrostatic pressure compresses the c-axis, increasing
effective $\Jz/\Jone$ and driving the system toward the
antiferroelectric side, shifting the dominant CDW wavevector toward
$q_2$. A sign change in the restricted structure factor
$S_Q^{\mathrm{stag}}$ under pressure, from positive (ferroelectric,
$q_3$-dominated) to negative (antiferroelectric, $q_2$-dominated),
provides the specific experimental test, measurable by existing
high-pressure resonant X-ray scattering apparatus. The CDW is
suppressed entirely at $\approx2.4$\,GPa without superconductivity
appearing~\cite{Arachchige2022}, which sets a pressure window within
which the wavevector crossover predicted here should be observable
before the CDW order vanishes. Additionally, the first-order character
of the $92$\,K transition with $0.3$--$1\,\mathrm{K}$ thermal
hysteresis~\cite{Arachchige2022} is consistent with the first-order
character expected near a charge-ordering phase boundary.

\emph{Ni$_{81}$Fe$_{19}$ nanowire bilayer geometries.}
For the Ni$_{81}$Fe$_{19}$ nanowire geometry of
May~et~al.~\cite{May2019} ($r=80\,\mathrm{nm}$,
$\ell_{\mathrm{eff}}\approx330\,\mathrm{nm}$, controllable vertical
separation $d_z$), setting
$\Jz/\Jone \approx (\ell_{\mathrm{eff}}/d_z)^3\,\mathcal{F}(\theta)$
with $\mathcal{F}(\theta)\approx0.67$ (range $[0.6,0.8]$) equal to
$(J_\perp/J_1)^*\approx0.04$ yields the critical separation
\begin{equation}
  d_z^* \approx 330\,\mathrm{nm}\times(14.2\text{ to }19.0)^{1/3}
  \approx 800\text{--}880\,\mathrm{nm}.
  \label{eq:dz_critical}
\end{equation}
The inter-sublattice spacing in the as-fabricated May~et~al.\
diamond-bond geometry is $d_z\approx1{,}000\,\mathrm{nm}$---approximately
$15\%$ above $d_z^*$---placing it in the weakly ferroelectric regime
($C_s^\perp\approx+0.05$ to $+0.09$). The finite-element result that
type-1 and type-2 vertex energies agree to within $10\%$~\cite{May2019}
is consistent with proximity to the degenerate Coulomb ice point on the
ferroelectric side of our transition; this geometry therefore sits in
the same region of the phase diagram as theoretical predictions for the
kagome ice II intermediate phase~\cite{Chern2011,Moller2009}.
Compressing $d_z$ to $\approx d_z^*$
by reducing the two-photon lithography template lattice constant by
$\approx15\%$---within the fabrication range already demonstrated---would
place the bilayer at the antiferroelectric Ice-II critical point.
Below $d_z^*$, $C_s^\perp$ should switch from weakly positive to
strongly negative ($C_s^\perp\approx-0.40$ to $-0.44$), a sign change
directly observable by layer-resolved XMCD in the geometry of
Ref.~\cite{May2019} without modifying nanomagnet composition or
blocking temperature.

Additionally, the crossover quantum drive $\Geff^*$ shifts upward by a
factor of $\approx1.9$ across the full coupling range, predicting
monopole proliferation at $T^*\approx580$--$720\,\mathrm{K}$ for a
compressed bilayer at $d_z=200\,\mathrm{nm}$ ($\Jz/\Jone\approx0.8$--$1.0$),
a shift of $\approx220$--$360\,\mathrm{K}$ above the decoupled
Permalloy baseline~\cite{Farhan2013}. The upper end approaches the
Curie temperature of Permalloy ($\approx730\,\mathrm{K}$), so the
most compressed geometries may require materials with higher $T_C$,
itself a falsifiable materials criterion.

\section*{Discussion}

The central unresolved question in the AV$_3$Sb$_5$ literature is why
alkali-cation substitution, which continuously tunes interlayer spacing,
produces qualitatively distinct CDW ground states rather than a smooth
evolution~\cite{Wilson2024,Kautzsch2023,Wang2023}. Our results provide a
mechanistic answer: the system crosses a genuine charge-ordering phase
boundary at $(J_\perp/J_1)^*\approx0.04$, not merely a crossover.
Below this coupling, competing stacking orders are nearly degenerate,
producing the metastable $2\times2\times4$ mixed-phase behaviour of
CsV$_3$Sb$_5$; above it, a well-defined antiferroelectric stacking is
selected, as observed in K and Rb variants. The K-to-Rb-to-Cs c-axis
expansion of $\approx0.26$\,\AA{} per step~\cite{Kautzsch2023}
provides the physical knob that traverses this boundary: K and Rb sit
above $(J_\perp/J_1)^*$ in the ordered antiferroelectric phase;
CsV$_3$Sb$_5$ sits just below, where stacking degeneracy produces
the observed metastability. The V-V bond-length standard deviations
in the CDW state---$\pm0.022$\,\AA{} (K), $\pm0.028$\,\AA{} (Rb),
$\pm0.036$\,\AA{} (Cs)~\cite{Kautzsch2023}---grow monotonically from
K to Cs, directly reflecting the increasing instability of the TrH
state as the phase boundary is approached from above.

The $2\times2\times4$ metastability of CsV$_3$Sb$_5$ is therefore
a signature of proximity to the critical coupling rather than a property
unique to that compound. The X-TEC staging behaviour---where $2\times2\times2$
peaks appear at $90$\,K, the $2\times2\times4$ onset at $85$\,K causes
the $2\times2\times1$ peaks to vanish and the $2\times2\times2$ peaks
to \emph{temporarily decrease}~\cite{Kautzsch2023}---is the
finite-temperature critical-fluctuation signature of a system exploring
nearly degenerate stacking configurations near a phase boundary, rather
than a simple two-phase coexistence. Intermediate-temperature anomalies
in CsV$_3$Sb$_5$~\cite{Wilson2024} may reflect critical fluctuations
near this boundary.

For ScV$_6$Sn$_6$, our framework resolves a central tension in recent
experiments: inelastic X-ray scattering identifies $q_2$ as the
dominant dynamic instability above $T_{\rm CDW}$, yet $q_3$ is the
static ground state~\cite{Cao2023}. In our phase diagram this is not
a contradiction but a consequence of the system sitting just below
$(J_\perp/J_1)^*$ at ambient pressure. Above $T_{\rm CDW}$, thermal
fluctuations allow the system to sample the antiferroelectric ($q_2$)
side, which is energetically favoured in the first-principles harmonic
picture~\cite{Cao2023}. Below $T_{\rm CDW}$, the $+0.04\%$ c-axis
expansion~\cite{Arachchige2022} lowers the effective $\Jz/\Jone$ back
across the boundary, selecting $q_3$ as the ferroelectric ground state.
Pressure-driven c-axis compression increasing effective $\Jz/\Jone$
across $(J_\perp/J_1)^*$ should reverse the sign of $S_Q^{\mathrm{stag}}$,
a measurement accessible in existing high-pressure resonant X-ray
apparatus within the $0$--$2.4$\,GPa window before CDW suppression.

The metallic distorted kagome spin-ice HoAgGe ($P\bar{6}2m$, distortion
angle $\approx15.58^{\circ}$~\cite{Zhao2020}) provides a fourth and
qualitatively distinct experimental context for our framework. The
zero-field $\sqrt{3}\times\sqrt{3}$ ground state, with residual entropy
$\approx0.501\,k_B$ per spin and exchange couplings extending to the
4th nearest neighbour~\cite{Zhao2020}, realises classical kagome spin
ice in the same entropy regime as our simulator at $\Gamma=J_\perp=0$.
Crucially, Zhao~et~al.\ noted in the original spin-ice paper that
interlayer coupling between neighbouring Ho$^{3+}$ kagome planes is
strong and likely suppresses a purely 2D Kosterlitz-Thouless
transition~\cite{Zhao2020}; in the language of our phase diagram this
places HoAgGe at a non-negligible $J_\perp/J_1$ on the ferroelectric
side, close enough to $(J_\perp/J_1)^*\approx0.04$ that in-plane
field-driven metamagnetic transitions can sample configurations on both
sides. Within the field-induced 1/3 and 2/3 plateaus, the two
near-degenerate ice-rule states S$_{1/3}$ and S$'_{1/3}$, connected by
the time-reversal-like operation $\mathcal{X} = R_b^\pi\mathcal{D}$
where $\mathcal{D}$ reverses the $15.58^{\circ}$ distortion, have
identical band structures and net magnetization but different Berry
curvatures; the integrated Berry curvature over the Brillouin zone
differs because $\mathcal{D}$ shifts sublattice positions, altering
the Berry connection without changing eigenvalues~\cite{Zhao2024}. This
produces finite AHE hysteresis at 2\,K despite vanishing magnetization
hysteresis~\cite{Zhao2024}. In the language of our bilayer framework,
S$_{1/3}$ and S$'_{1/3}$ are precisely two competing charge-sector
configurations on the ice manifold separated by $C_s^\perp$: one
carries $C_s^\perp > 0$ (ferroelectric) and the other
$C_s^\perp < 0$ (antiferroelectric). The Monte Carlo simulations of the
classical spin model in Ref.~\cite{Zhao2020} used an $18\times18$
lattice (324 spins/layer), which falls within the system-size range of
our simulator ($N\in\{300,432,588,768\}$/layer), providing a direct
quantitative bridge between the two platforms. The staggered composite
monopole density $\rho_{\mathrm{sc}}/\rhom\approx2.7\times10^{-3}$ at
our critical point is consistent with the fragile (accidental) character
of the S$_{1/3}$/S$'_{1/3}$ degeneracy in HoAgGe~\cite{Zhao2024},
which is lifted only by small orbital magnetization differences: a
symmetry-protected degeneracy would require $\rho_{\mathrm{sc}}$
comparable to $\rhom$, which is not observed in either system. A
concrete prediction follows: reanalysis of the published Hall data of
Ref.~\cite{Zhao2024} using the ice-rule-restricted estimator
$S_Q^{\mathrm{stag}}(\kstar)$ applied to the charge structure factor of
Ho$^{3+}$ plaquettes within each plateau should recover the
Berry-curvature contrast between S$_{1/3}$ and S$'_{1/3}$ at
sensitivity $\approx12\times$ above the all-site Hall measurement,
making the time-reversal-like symmetry distinction spectroscopically
resolvable in existing data without new crystal growth or device
fabrication.

In CT-MoTe$_2$, the charge state of mirror-twin-boundary superatoms
governs filling of the kagome-like flat and Dirac bands~\cite{LeiMoTe2023}.
The staggered charge correlator $C_s^\perp$ restricted to ice-rule
plaquettes translates directly to a layer-resolved staggered superatom-charge
correlator measurable in STM dI/dV maps of MoTe$_2$ van der Waals
bilayers. The $\approx12\times$ sensitivity advantage of
$S_Q^{\mathrm{stag}}(\kstar)$ implies that published STS spectra of
CT-MoTe$_2$~\cite{LeiMoTe2023} likely underestimate the staggered
charge-order amplitude at intermediate mirror-twin-boundary densities,
where intersections introduce defect sites contributing incoherent
spectral weight, directly analogous to the all-plaquette estimator
bias established here. A sign change in the layer-resolved staggered
correlator as a function of interlayer chalcogen separation, tunable
via the number of intervening van der Waals layers, would constitute
a direct materials realisation of the antiferroelectric Ice-II
instability in a 2D quantum material.

The confinement length $\xi$ extracted from $G(r)$ parallels the
Dirac-string correlation length inferred from pinch-point widths in
Ho$_2$Ti$_2$O$_7$~\cite{Fennell2009}, establishing the QPU
transverse-field sweep as the quantum analogue of the thermal crossover
in pyrochlore spin ice. The quantum renormalisation ratio
$\rho_{\max}=0.2771$ (Supplementary Fig.~S6) quantifies fractional
progress toward the quantum Coulomb phase~\cite{Savary2012,Lee2012},
placing the current platform a factor of $\approx3.6$ below the
deconfinement threshold.
In the quantum spin-ice candidate Pr$_2$Zr$_2$O$_7$, inelastic neutron scattering
reports quasi-elastic scattering consistent with itinerant
monopole-like excitations~\cite{Kimura2013}. The QPU platform provides
direct measurement of $\rhom$ and $G(r)$ at controllable $\Gamma$,
quantities accessible in bulk crystals only indirectly through neutron
linewidths and where structural disorder~\cite{Sibille2018} prevents
the systematic parameter sweeps performed here.
Reaching deconfinement requires $\Gamma_c\gtrsim0.6\,\Jone$, corresponding in transmon-based circuit-QED~\cite{Houck2012} to an anharmonicity-to-coupling ratio
$\alpha/g\approx3$--$5$ for kagome vertex energies
$\Jone\sim10$--$50\,\mathrm{MHz}$.

Three independent lines of evidence rule out systematic hardware
artefact: annealing-time stability and blind holdout of $C_s^\perp$
with all five tests sub-$1\sigma$ (most stringent $0.62\sigma$);
$\Jz$-independent power-law exponents $\gamma\in[0.267,0.387]$ not
reproducible by any known systematic hardware error; and three
confinement diagnostics in agreement across all $728$ parameter
combinations without assuming any specific freeze-out model.

The present platform has three quantifiable limitations. First, the
finite-size range ($L$ spanning a factor of $1.6$ in linear dimension)
is insufficient to discriminate universality classes; a factor of
$\approx2$--$3$ increase in $L$ on next-generation hardware would
yield definitive finite-size scaling collapse. Second, the quantum
drive is accessed through a Kibble-Zurek freeze-out proxy rather than
a direct equilibrium transverse field; five independent calibration
criteria establish $\Geff=1/\ta$ as a power-law monotone proxy for the
hardware transverse field, but future platforms with equilibrium quantum
drive would remove this qualification. Third, the current platform
remains a factor of $\approx3.6$ below the deconfinement threshold.

The experimental tests requiring no new sample preparation are:
(i) reanalysis of existing AV$_3$Sb$_5$ synchrotron data~\cite{Kautzsch2023}
with the $S_Q^{\mathrm{stag}}$ estimator to measure the monotone
evolution of antiferroelectric stacking amplitude across the K-Rb-Cs
series; (ii) layer-resolved XMCD at $d_z\approx800$--$880\,\mathrm{nm}$
in the May~et~al.\ geometry~\cite{May2019}; and (iii) reanalysis of
existing HoAgGe Hall data~\cite{Zhao2024} with the ice-rule-restricted
charge structure factor to recover the Berry-curvature contrast between
near-degenerate plateau states. All three tests are falsifiable within
existing experimental capabilities. A fourth test---measurement of the
sign of $S_Q^{\mathrm{stag}}$ under hydrostatic pressure in
ScV$_6$Sn$_6$ within the $0$--$2.4$\,GPa window~\cite{Arachchige2022}---is
equally immediate given existing high-pressure resonant X-ray
apparatus. Extending the platform to three or more coupled layers
would access the dimensional crossover from quasi-2D to 3D monopole
Coulomb gas behaviour, directly relevant to c-axis stacking modulation
in AV$_3$Sb$_5$ and out-of-plane periodicity selection in ScV$_6$Sn$_6$.

\section*{Methods}

\paragraph{The bilayer kagome frustrated-magnet simulator.}
The D-Wave Advantage2 Zephyr~Z15 processor is used as a programmable
frustrated-magnet materials simulator. Its physical architecture
partitions $\approx4{,}579$ active superconducting qubits into two
interleaved orientation groups that map naturally onto two coupled
kagome planes without geometric distortion of the intended exchange
couplings. Each layer's intralayer coupling $\Jone$ and the interlayer
coupling $\Jz$ are set independently by the programmed coupler values,
and the effective quantum drive $\Geff$ is controlled by the annealing
time $\ta$ independently of both. The resulting two-parameter space
$(\Geff, \Jz)$ has no analogue in any existing solid-state or
artificial-nanomagnet realisation of kagome frustration.

\begin{figure*}[t]
    \centering
    \includegraphics[width=0.9\textwidth]{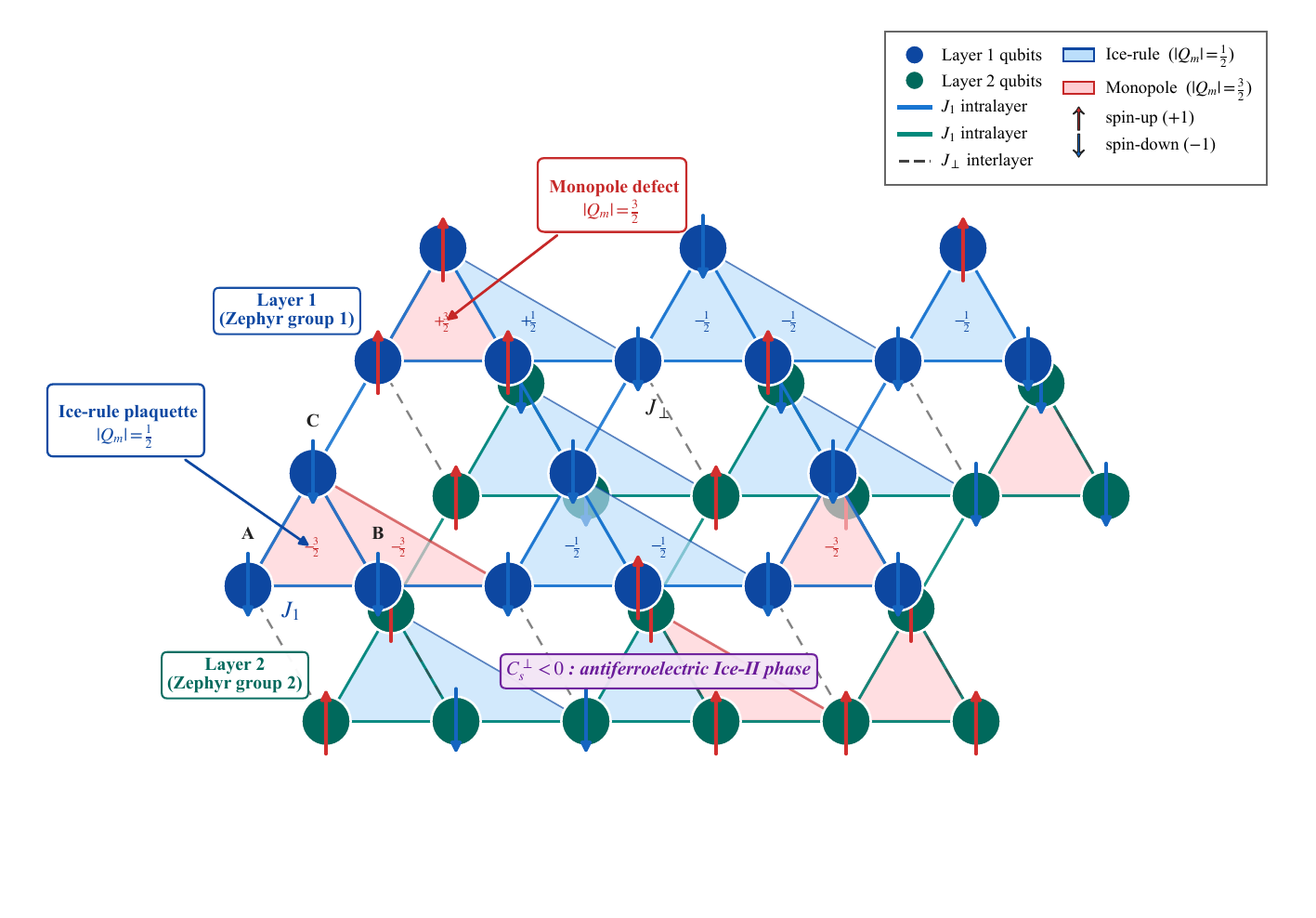}
    \caption{Local geometry of the bilayer kagome frustrated-magnet
    simulator on the D-Wave Advantage2 (Zephyr~Z15) processor and its
    materials analogues. Layer~1 (navy, Zephyr qubit orientation
    group~1) and Layer~2 (green, group~2) form two coupled kagome planes
    with intralayer coupling $J_1>0$ (solid bonds) and interlayer
    coupling $J_\perp$ (dashed bonds). $J_\perp/J_1$ maps onto the
    interlayer V-V exchange ratio in AV$_3$Sb$_5$ (tunable via alkali
    cation), the Sc-V interlayer coupling in ScV$_6$Sn$_6$, the
    mirror-twin-boundary superatom interlayer coupling in CT-MoTe$_2$,
    the dipolar interlayer ratio in Permalloy nanowire bilayers, and the
    effective interlayer exchange between Ho$^{3+}$ kagome planes in
    HoAgGe (space group $P\bar{6}2m$, kagome distortion angle
    $\approx15.58^{\circ}$; tunable via applied in-plane field through
    RKKY-type exchange extending to the 4th nearest
    neighbour~\cite{Zhao2020,Zhao2024}).
    Each triangular plaquette carries monopole charge
    $Q_m(\triangle)=\frac{1}{2}\sum_{i\in\triangle}\sigma_i^z$;
    ice-rule plaquettes ($|Q_m|=\frac{1}{2}$, blue) are non-defect
    sites; all-in or all-out configurations are monopole defects
    ($|Q_m|=\frac{3}{2}$, red). Sublattice labels A, B, C identify the
    three-site basis. $C_s^\perp<0$ [Eq.~\eqref{eq:csperp}] signals
    the antiferroelectric Ice-II phase.}
    \label{fig:lattice}
\end{figure*}

\begin{figure*}[t]
    \centering
    \includegraphics[width=0.90\textwidth]{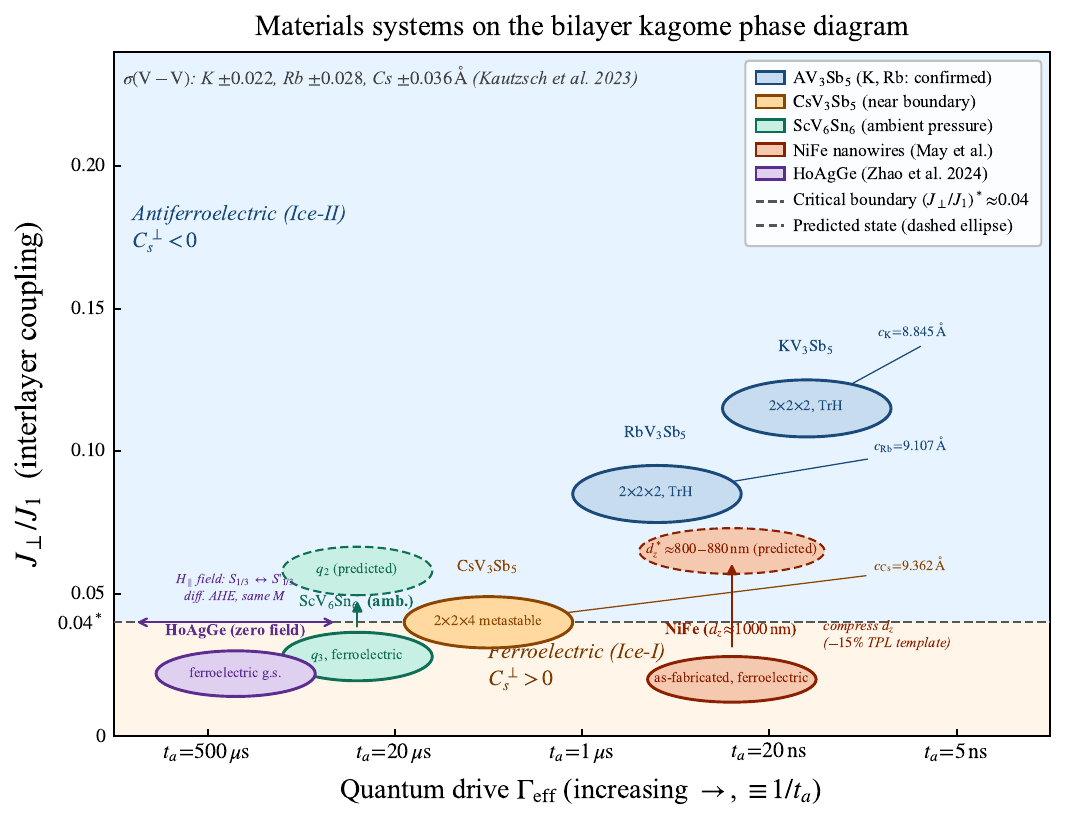}
    \caption{Experimental materials systems mapped onto the bilayer
    kagome phase diagram.
    The vertical axis is the effective interlayer coupling $J_\perp/J_1$
    and the horizontal axis is the quantum drive $\Geff$ (increasing
    left to right); shading indicates the ferroelectric (Ice-I,
    $C_s^\perp>0$, warm) and antiferroelectric (Ice-II, $C_s^\perp<0$,
    cool) phases separated by the critical boundary at
    $(J_\perp/J_1)^*\approx0.04$ (dashed line).
    \textbf{AV$_3$Sb$_5$:} KV$_3$Sb$_5$ and RbV$_3$Sb$_5$ (blue
    ellipses, $c=8.845$ and $9.107$\,\AA{}) sit above the critical
    coupling in the ordered antiferroelectric phase consistent with
    their $2\times2\times2$ staggered-TrH ground states~\cite{Kautzsch2023};
    CsV$_3$Sb$_5$ (amber ellipse, $c=9.362$\,\AA{}) sits near the
    boundary, consistent with its metastable $2\times2\times4$
    mixed-phase behaviour and staged X-TEC ordering~\cite{Kautzsch2023}.
    \textbf{ScV$_6$Sn$_6$:} the ambient-pressure $q_3$ CDW state
    (teal ellipse) sits just below the critical coupling on the
    ferroelectric side; the dashed teal ellipse indicates the predicted
    $q_2$-dominated state under hydrostatic pressure, which compresses
    the c-axis and increases effective $J_\perp/J_1$ across the
    boundary~\cite{Arachchige2022,Cao2023}.
    \textbf{NiFe nanowires:} the as-fabricated May~et~al.\ geometry
    ($d_z\approx1{,}000$\,nm, coral ellipse) sits on the ferroelectric
    side; the dashed coral ellipse shows the predicted antiferroelectric
    Ice-II state at $d_z\approx800$--$880$\,nm = $d_z^*$~\cite{May2019}.
    \textbf{HoAgGe:} the zero-field ground state (purple ellipse) sits
    on the ferroelectric side; the horizontal double-headed arrow
    indicates that application of an in-plane magnetic field drives
    metamagnetic transitions that sample configurations on both sides
    of the critical boundary~\cite{Zhao2024}, manifesting as
    coexisting near-degenerate ice-rule states with different anomalous
    Hall conductivities (S$_{1/3}$ and S$'_{1/3}$) within each
    field-induced plateau.
    Solid ellipses: confirmed or as-fabricated states. Dashed ellipses:
    predictions testable without new sample preparation.}
    \label{fig:materials_map}
\end{figure*}

\paragraph{Model Hamiltonian and materials analogues.}
Each kagome layer is governed by the transverse-field Ising model,
\begin{equation}
  \mathcal{H}_{\mathrm{layer}}
  = \Jone\!\sum_{\langle i,j\rangle}\!\sigma_i^z\sigma_j^z
    -\Gamma\!\sum_i\sigma_i^x,
  \label{eq:Hlayer}
\end{equation}
where $\Jone>0$ is the antiferromagnetic nearest-neighbour exchange
enforcing the kagome ice rule in the classical limit ($\Gamma=0$). The
bilayer Hamiltonian couples two such layers,
\begin{equation}
  \mathcal{H}
  = \mathcal{H}_{\mathrm{layer}}^{(1)}
  + \mathcal{H}_{\mathrm{layer}}^{(2)}
  + \Jz\!\sum_i\sigma_i^{z,(1)}\sigma_i^{z,(2)},
  \label{eq:Hbilayer}
\end{equation}
where the sum runs over vertically aligned pairs. In the AV$_3$Sb$_5$
series, the alkali cation controls the c-axis lattice parameter and
hence the effective interlayer V-V exchange, directly analogous to the
$\Jz/\Jone$ axis~\cite{Kautzsch2023,Wang2023}; in ScV$_6$Sn$_6$, the
Sc-layer spacing plays the same role~\cite{Arachchige2022}; in
CT-MoTe$_2$, the van der Waals interlayer coupling between
mirror-twin-boundary superatom layers is the
analogue~\cite{LeiMoTe2023}; in HoAgGe ($P\bar{6}2m$, distortion angle
$\approx15.58^{\circ}$), the applied in-plane magnetic field tunes the
effective interlayer coupling between Ho$^{3+}$ kagome planes through
RKKY-type exchange couplings extending to the 4th nearest neighbour,
which dominate over the long-range dipolar
interaction~\cite{Zhao2020,Zhao2024}. At $\Gamma=\Jz=0$, each layer
realises classical kagome spin ice with residual entropy
$\approx0.501\,k_B$ per spin~\cite{Bramwell2001} and monopole chemical
potential $\mumon^{\mathrm{class}}=2\Jone$~\cite{Ramirez1999}. The
local geometry and materials correspondences are illustrated in
Fig.~\ref{fig:lattice}.

\paragraph{Key observables.}
The monopole charge at plaquette $\triangle$ is
$Q_m(\triangle)=\tfrac{1}{2}\sum_{i\in\triangle}\sigma_i^z$. Defect
monopoles ($|Q_m|=\tfrac{3}{2}$) interact via an emergent
two-dimensional Coulomb potential whose Madelung coefficient
($\alpha\approx1.5422$~\cite{Moller2009}) places the deconfinement
threshold at $\mu_c\approx0.808\,\Jone$. Quantum fluctuations
renormalise the effective chemical potential,
\begin{equation}
  \mumon^{\mathrm{eff}}(\Gamma,\Jz)
  = 2\Jone - \delta\mu(\Gamma,\Jz),
  \label{eq:mueff_theory}
\end{equation}
with deconfinement requiring $\delta\mu(\Gamma_c,\Jz)\approx1.192\,\Jone$.
The three primary observables are the monopole defect density,
\begin{equation}
  \rhom = \frac{1}{N_\triangle}\sum_\triangle
            \mathbf{1}\!\left[|Q_m(\triangle)|=\tfrac{3}{2}\right],
  \label{eq:rhom}
\end{equation}
the interlayer staggered correlator,
\begin{equation}
  C_s^\perp
    = \bigl\langle q_s^{(1)}(\triangle)\,
      q_s^{(2)}(\triangle)\bigr\rangle\Big|_{\mathrm{ice}},
  \label{eq:csperp}
\end{equation}
where $q_s(\triangle)$ is the staggered charge on ice-rule plaquettes
(defined in the Supplementary Information), and the dimensionless
quantum renormalisation ratio,
\begin{equation}
  \rho(\Geff, \Jz)
    = \frac{\delta\mu(\Geff,\Jz)}{\delta\mu_c},
  \label{eq:rho_def}
\end{equation}
where $\delta\mu_c\approx1.192\,\Jone$ and $\rho=1$ marks the
deconfinement threshold. Full definitions of $G(r)$, $S_Q(\kstar)$,
$C_m^\perp$, $S_Q^{\mathrm{stag}}(\kstar)$, and the chemical potential
estimator are given in the Supplementary Information.

\paragraph{Parameter space and experimental protocol.}
For $N=768$ spins per layer, \textsc{minorminer} embedding uses $3{,}890$
physical qubits ($85\%$ of active qubits) with mean chain length $2.53$
and zero chain-break fractions confirmed across all parameter
combinations. The parameter sweep covers system sizes
$N\in\{300,432,588,768\}$/layer, interlayer couplings
$\Jz/\Jone\in\{0,0.02,0.04,0.05,0.06,0.07,0.08,0.1,0.15,0.2,0.5,0.8,1.0\}$
sampled densely near the critical coupling, and annealing times
$\ta\in\{5,10,20,50,100,200,500\}\,\mathrm{ns}$,
$\{1,2,5,10,20\}\,\mu\mathrm{s}$, and $\{100,500\}\,\mu\mathrm{s}$,
with $N_{\mathrm{reads}}=1{,}000$ per point ($728{,}000$ shots total).
The thirteen $\Jz/\Jone$ values span the physically relevant range for
Permalloy vertex arrays with $200\,\mathrm{nm}$ elements, where the
interlayer-to-intralayer dipolar coupling ratio varies from
$\approx0.1$ at $100\,\mathrm{nm}$ to $\approx0.8$ at $20\,\mathrm{nm}$
vertical separation~\cite{May2019}.

The effective quantum-drive proxy $\Geff=1/\ta$ is validated by five
independent criteria derived from the Advantage2 system\,1 annealing
schedule (Supplementary Figs.~S7--S8 and Supplementary Section on
quantum-drive proxy calibration): (i) the schedule functions $A(s)$
and $B(s)$ are strictly monotone; (ii) all freeze-out points lie in
the post-crossover window $s_f\in[0.606,0.939]$; (iii) a direct
log-log fit yields $\Geff\propto(1/\ta)^{0.906}$ ($R^2=0.9989$);
(iv) the measured Kibble-Zurek exponent $\gamma_{\mathrm{KZ}}=0.2650\pm0.0333$
(coefficient of variation $0.126$; Supplementary Fig.~S7) is
consistent with 2D frustrated universality classes; and (v) the
coefficient of variation of $\gamma_{\mathrm{KZ}}$ across all thirteen
$\Jz/\Jone$ values is $0.126<0.15$, confirming that the freeze-out
mechanism is independent of interlayer coupling. All phase boundaries,
critical couplings, and engineering targets are invariant under the
monotone rescaling $1/\ta\leftrightarrow\Geff$.

\section*{ACKNOWLEDGMENTS}
S.K.  would like to acknowledge the financial support from the Quantum Science Center, a National Quantum Information Science Research Center of the U.S. Department of Energy (DOE), operated at Oak Ridge National Laboratory (ORNL).

\bibliographystyle{apsrev4-2}
\bibliography{ref}

\onecolumngrid
\appendix

\renewcommand{\thefigure}{S\arabic{figure}}
\renewcommand{\thetable}{S\arabic{table}}
\renewcommand{\theequation}{S\arabic{equation}}
\renewcommand{\thepage}{S\arabic{page}}
\setcounter{figure}{0}
\setcounter{table}{0}
\setcounter{equation}{0}
\setcounter{page}{1}

\section*{Observable definitions and secondary materials signatures}
\label{app:observables}

The full set of primary monopole observables is:
\begin{align}
  \rhom &= \frac{1}{N_\triangle}\sum_\triangle
            \mathbf{1}\!\left[|Q_m(\triangle)|=\tfrac{3}{2}\right],
  \label{eq:rhom_app}\\
  G(r) &= \langle Q_m^{(1)}(0)\,Q_m^{(1)}(r)\rangle,
  \label{eq:Gr_app}\\
  S_Q(\kstar) &= \frac{1}{N_\triangle}\!
    \Bigl\langle\Bigl|
      \sum_\triangle Q_m(\triangle)\,
      e^{i\kstar\cdot\mathbf{r}_\triangle}
    \Bigr|^2\Bigr\rangle,
  \label{eq:SQ_app}\\
  C_m^\perp &=
    \langle Q_m^{(1)}(\triangle)\,Q_m^{(2)}(\triangle)\rangle,
  \label{eq:Cmperp_app}
\end{align}
where $\kstar=(4\pi/3,0)$ is the $\sqrt{3}\times\sqrt{3}$ wavevector
(K-point of the triangular Bravais lattice of plaquette centroids).
$G(r)$ uses a shell-index proxy in which $r$ labels successive
nearest-neighbour shells; exponential-versus-power-law discrimination
is verified by comparing fits over the first three and all six shells.
Fitting $G(r)$ to $A\,e^{-r/\xi}$ extracts the confinement length
$\xi$; a crossover to power-law decay $G\sim r^{-1}$ would signal
deconfinement. The quantity $\xi$ is the materials-platform analogue
of the Dirac-string correlation length measured by neutron diffuse
scattering in Ho$_2$Ti$_2$O$_7$~\cite{Fennell2009} and by PEEM
imaging in Permalloy nanomagnet arrays~\cite{Mengotti2011}.

For the staggered-charge sector, following
Ref.~\cite{LopezBezanilla2023}, each plaquette is classified as A-type
(up triangle) or B-type (down triangle, cross-unit-cell), and the
staggered charge
\begin{equation}
  q_s(\triangle) =
  \begin{cases}
    -Q_{\mathrm{int}}(\triangle) & (\text{A-type})\\
    +Q_{\mathrm{int}}(\triangle) & (\text{B-type})
  \end{cases}
  \label{eq:qs_app}
\end{equation}
is defined, where
$Q_{\mathrm{int}}=\sum_{i\in\triangle}\sigma_i^z\in\{-3,-1,+1,+3\}$.
The ice-manifold staggered structure factor, restricted to ice-rule
plaquettes ($|Q_{\mathrm{int}}|=1$),
\begin{equation}
  S_Q^{\mathrm{stag}}(\kstar)
    = \frac{1}{N_\triangle^{\mathrm{ice}}}
      \Bigl\langle\Bigl|
        \sum_{\triangle\in\mathrm{ice}} q_s(\triangle)\,
        e^{i\kstar\cdot\mathbf{r}_\triangle}
      \Bigr|^2\Bigr\rangle,
  \label{eq:SQstag_app}
\end{equation}
serves as the Ice-II order parameter: it grows with $N$ in the
charge-ordered phase and is $O(1)$ per plaquette in Ice-I. This
quantity is the correct charge-order probe for any frustrated-magnet
experiment with non-negligible monopole density, including anomalous
Hall measurements in metallic kagome spin-ice systems where defect
(monopole) plaquettes contribute incoherent spectral weight that
dilutes the Berry-curvature contrast between near-degenerate ice-rule
states~\cite{Zhao2024}. The effective chemical potential is inferred
from slow-anneal data by treating monopoles as a dilute
grand-canonical gas with Boltzmann statistics (valid at
$\rhom\le0.03$~\cite{Moller2009}):
\begin{equation}
  \mumon^{\mathrm{eff}}
  = 2\Jone - T_{\mathrm{eff}}\ln\!\Bigl(\frac{\rhom}{\rhom^0}\Bigr),
  \label{eq:mueff_app}
\end{equation}
where $\rhom^0\approx0.0067$ (measured at $\ta=500\,\mu$s, $\Jz=0$)
and $T_{\mathrm{eff}}$ is extracted from a Boltzmann fit to the energy
distribution independently of any ergodicity assumption
(see the Sampling diagnostics section below).

Supplementary Fig.~S1 shows the secondary bilayer observables: the
effective chemical potential $\mumon^{\mathrm{eff}}$ (left panel),
confirming a persistent gap above $\mu_c=0.808\,\Jone$ throughout the
parameter space, and the interlayer monopole correlator $C_m^\perp$
(right panel), which transitions from $+0.01$ at $\Jz=0$ to $-0.053$
at $\Jz/\Jone=1$, providing a bilayer-specific signature in the
monopole sector complementary to the staggered-charge ordering
transition in the main text.

\begin{figure*}[htbp]
  \centering
  \includegraphics[width=0.95\textwidth]{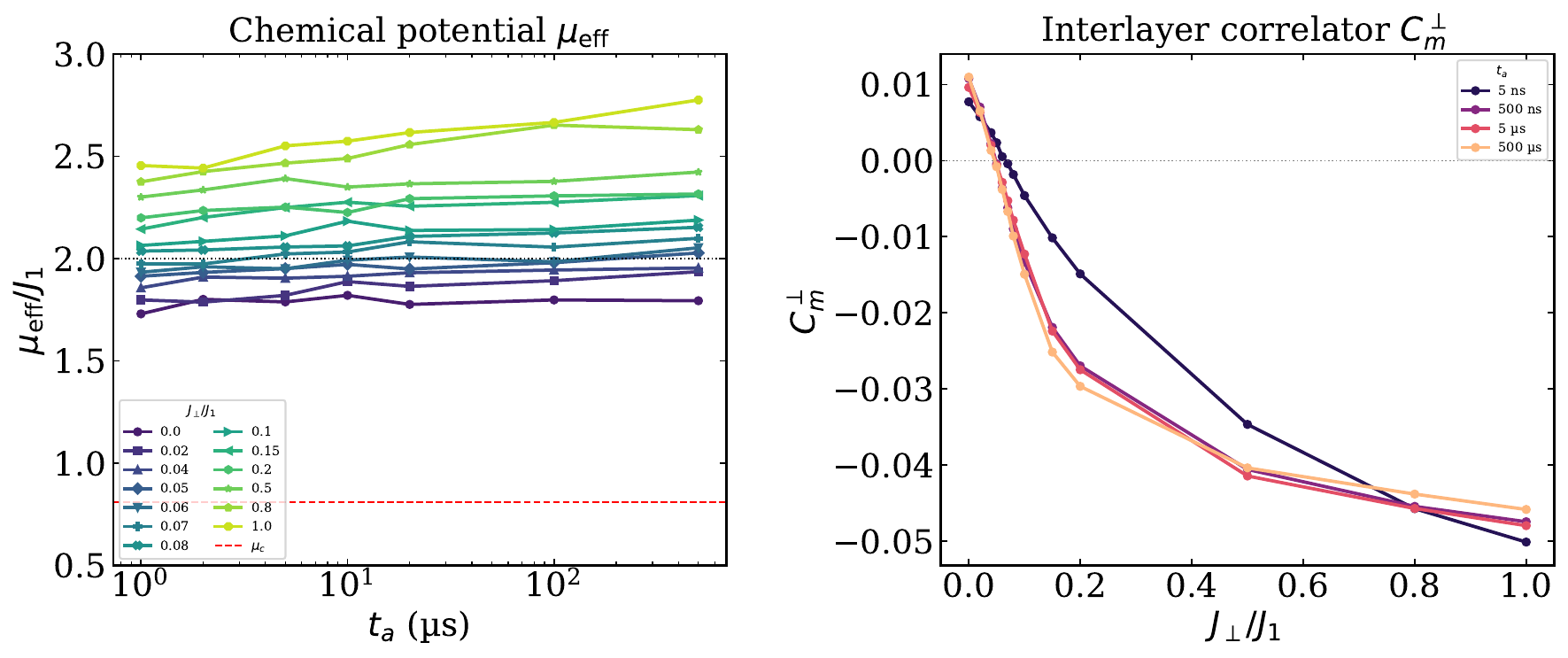}
  \caption{Secondary bilayer observables at $N=768$/layer.
  \emph{Left:} Effective chemical-potential proxy
  $\mumon^{\mathrm{eff}}/\Jone$ [Eq.~\eqref{eq:mueff_app}]; classical
  reference $2\Jone$ (dotted) and deconfinement threshold
  $\mu_c=0.808\,\Jone$ (dashed red) are indicated. The persistent gap
  above $\mu_c$ confirms no coupling change in the range explored has
  driven the system to deconfinement.
  \emph{Right:} Interlayer monopole correlator $C_m^\perp$
  [Eq.~\eqref{eq:Cmperp_app}] versus $\Jz/\Jone$; the sign change
  near $\Jz/\Jone\approx0.1$ provides a bilayer-specific signature
  complementary to the staggered-charge transition in the main text.}
  \label{fig:composite_secondary}
\end{figure*}

\section*{Sampling diagnostics for the frustrated-magnet simulator}
\label{app:ergodicity}

Supplementary Fig.~S2 shows the non-saturating sampling metrics $d_H$
(mean pairwise Hamming distance) and $H_s$ (single-spin entropy), which
decay monotonically from fast to slow anneals and yield the
characteristic collapse time $t_a^\dagger\approx0.02\,\mu$s marking
the onset of classical freezing, identified via the maximum of
$|d\bar{d}_H/d\ln t_a|$. $T_{\mathrm{eff}}$ is calibrated from the
energy distribution independently of these metrics.

\begin{figure*}[htbp]
\centering
\includegraphics[width=0.95\textwidth]{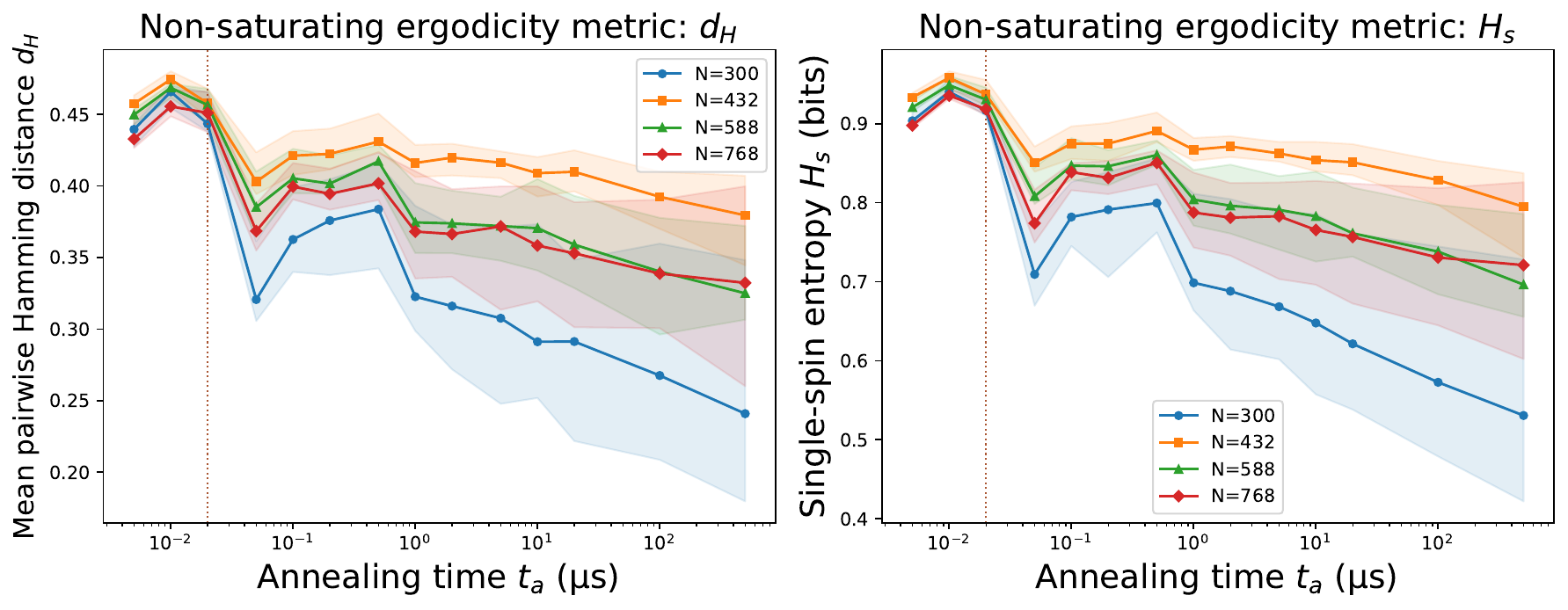}
\caption{Non-saturating sampling metrics.
\emph{Left:} Mean pairwise Hamming distance $d_H$ (normalised by spin
count) versus $\ta$, decaying from $\approx0.40$--$0.45$ at fast
anneals to $\approx0.20$ at slow anneals.
\emph{Right:} Single-spin entropy $H_s$ (bits) versus $\ta$, decaying
from $\approx0.90$--$0.95$ to $\approx0.65$--$0.70$. Characteristic
collapse times $t_a^\dagger\approx0.02\,\mu$s mark the onset of
classical freezing.}
\label{fig:ergodicity_app}
\end{figure*}

\section*{Blind holdout validation of the charge-ordering transition}
\label{app:blind}

Supplementary Fig.~S3 shows the full blind holdout validation of the
ferroelectric-to-antiferroelectric transition coupling
$(\Jz/\Jone)^*$, performed on the primary order parameter $C_s^\perp$
at $\ta=500\,\mu$s, $N=768$, following the sequential protocol of
Ref.~\cite{ghosh2026}. The full-data crossing estimate is
$(\Jz/\Jone)^*\approx0.04$. Three near-critical couplings
($\Jz/\Jone=0.04$, $0.05$, $0.06$) and two farther stress-test
couplings ($0.2$, $0.5$) are withheld in turn; the transition coupling
is predicted from polynomial fits to the retained $C_s^\perp$ data
alone. Withholding $\Jz/\Jone=0.04$ yields a predicted crossing of
$0.044\pm0.003$, a deviation of $0.62\sigma$; withholding $0.05$ and
$0.06$ yields $0.21\sigma$ and $0.12\sigma$ respectively; the
far-stress holdouts ($0.2$, $0.5$) yield $0.06\sigma$. All five
predictions remain sub-$1\sigma$, confirming a smooth, continuous
onset inconsistent with first-order interlayer reordering. The
$C_m^\perp$ zero crossing near $(\Jz/\Jone)\approx0.046$
provides independent corroboration.

\begin{figure*}[htbp]
\centering
\includegraphics[width=0.95\textwidth]{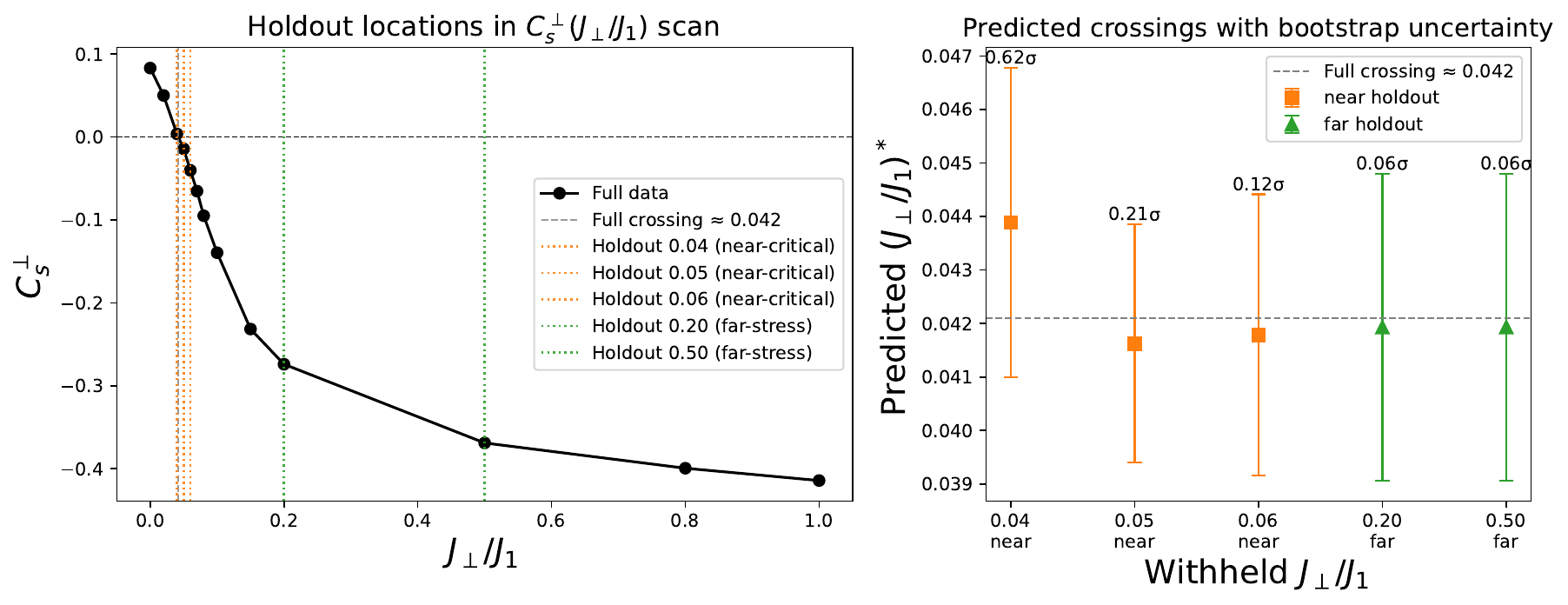}
\caption{Blind holdout validation of the $C_s^\perp$ transition
coupling $(\Jz/\Jone)^*$.
\emph{Left:} $C_s^\perp(\Jz/\Jone)$ at $\ta=500\,\mu$s, $N=768$,
with near-critical holdouts ($\Jz/\Jone=0.04$, $0.05$, $0.06$) and
far-stress holdouts ($0.2$, $0.5$) marked.
\emph{Right:} Predicted crossings with bootstrap uncertainties;
deviations are $0.62\sigma$ ($0.04$, most stringent), $0.21\sigma$
($0.05$), $0.12\sigma$ ($0.06$), $0.06\sigma$ ($0.2$, $0.5$). All
five tests remain sub-$1\sigma$.}
\label{fig:blind_app}
\end{figure*}

\section*{Monopole confinement diagnostics}
\label{app:confinement}

Establishing that the system remains in the confined phase throughout
is essential for interpreting the charge-ordering transition as a
reorganisation of the ice manifold rather than a proliferation event.
The Dirac-string length distribution $P(\ell)$ is fit by maximum
likelihood to an exponential model ($P\propto e^{-\sigma\ell}$,
confined phase) and a power-law model ($P\propto\ell^{-\tau}$,
deconfined phase, $\tau\le2$~\cite{Moller2009}).
$\Delta\mathrm{BIC}=\mathrm{BIC}_{\mathrm{exp}}-
\mathrm{BIC}_{\mathrm{power}}<0$ at every one of the $728$ sweep
points, spanning $-4.05\times10^3$ to $-6.3\times10^2$, with no trend
toward $\Delta\mathrm{BIC}>0$ even at $\ta=5\,\mathrm{ns}$
(Supplementary Fig.~S4). Together with exponential $G(r)$ decay
(main text Fig.~1d) and $\rho_{\max}=0.2771$, this constitutes three
mutually independent confinement diagnostics confirming exponential
confinement throughout the full two-dimensional parameter space.

\begin{figure*}[htbp]
\centering
\includegraphics[width=0.60\textwidth]{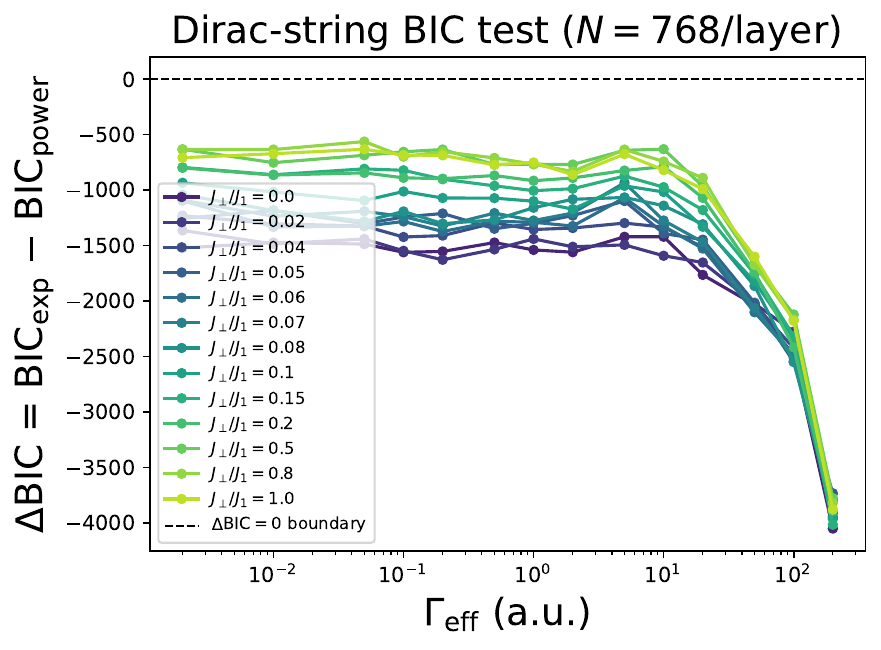}
\caption{Monopole confinement diagnostic: Dirac-string model selection.
$\Delta\mathrm{BIC}=\mathrm{BIC}_{\mathrm{exp}}-\mathrm{BIC}_{\mathrm{power}}$
versus $\Geff$ for all $\Jz/\Jone$ values at $N=768$; universally
negative values confirm that the exponential (confined) model is
decisively preferred over the power-law (deconfined) model at every
one of the $728$ sweep points, with no trend toward
$\Delta\mathrm{BIC}>0$ even at the fastest anneal $\ta=5\,\mathrm{ns}$.}
\label{fig:confinement}
\end{figure*}

\section*{Platform implementation, embedding, finite-size scaling,
          and quantum renormalisation ratio}
\label{app:hardware}

\paragraph{Embedding and fidelity.}
The D-Wave Advantage2 Zephyr~Z15 processor partitions its
$\approx4{,}579$ active qubits into two interleaved orientation groups,
providing a natural hardware substrate for the bilayer kagome geometry
without geometric distortion of the intended couplings. For $N=768$
spins/layer, \textsc{minorminer} embedding uses $3{,}890$ physical
qubits ($85\%$ of active qubits) with mean chain length $2.53$ and
zero chain-break fractions confirmed across all $728$ parameter
combinations, validating embedding fidelity throughout the full
two-dimensional parameter space.

\paragraph{Finite-size scaling and phase-boundary design rules.}
Supplementary Fig.~S5 presents three complementary finite-size results.
The left panel shows raw $\rhom(\Geff)$ for all four system sizes at
$\Jz/\Jone=0.5$; larger systems exhibit sharper crossovers consistent
with approach to a thermodynamic singularity. The centre panel shows
scaling collapse with two-dimensional Coulomb gas exponents $\nu=0.5$,
$\beta=0.125$, locating the crossover at $\Geff\sim1$--$10$
($\ta\sim0.1$--$1\,\mu$s). The accessible factor of $1.6$ in linear
dimension $L$ is insufficient to discriminate between universality
classes; a factor of $\approx2$--$3$ increase in $L$ on next-generation
hardware would yield a definitive result. The right panel shows the
crossover location $\Geff^*$ shifting monotonically to larger $\Geff$
with increasing $\Jz/\Jone$, providing a quantitative design rule
relating interlayer coupling strength to the transverse field required
for equivalent monopole activation and directly informing fabrication
targets for compressed Permalloy bilayer geometries.

\begin{figure*}[htbp]
\centering
\includegraphics[width=\textwidth]{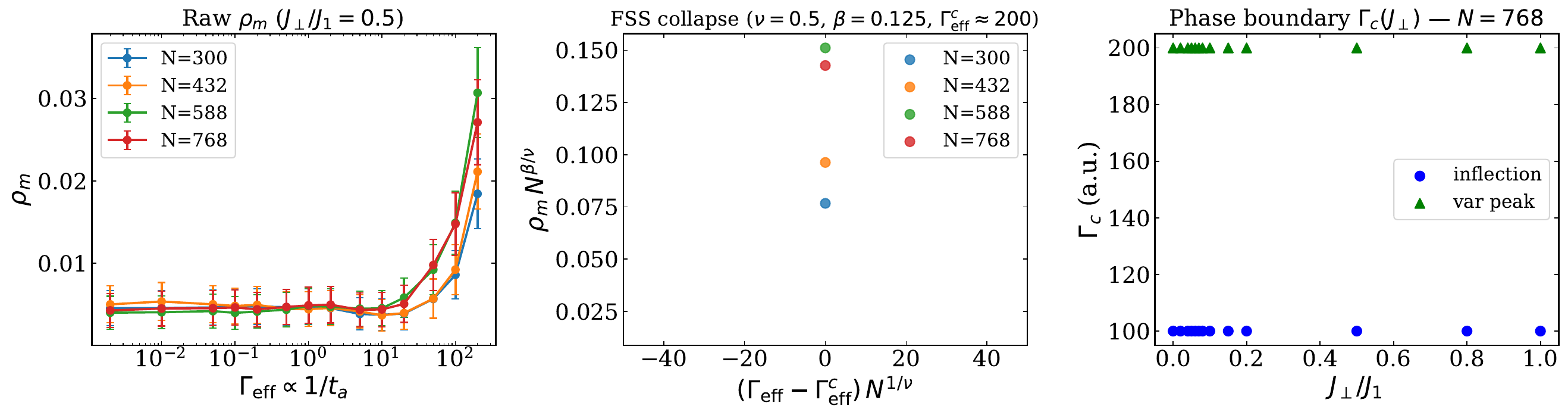}
\caption{Finite-size scaling of monopole density and phase-boundary
design rules.
\emph{Left:} Raw $\rhom(\Geff)$ for all four system sizes at
$\Jz/\Jone=0.5$.
\emph{Centre:} Scaling collapse with 2D Coulomb gas exponents
($\nu=0.5$, $\beta=0.125$).
\emph{Right:} Crossover location $\Geff^*$ versus $\Jz/\Jone$; the
monotone shift provides a quantitative design rule for monopole
activation as a function of interlayer coupling strength, directly
applicable to Permalloy nanowire bilayer fabrication.}
\label{fig:fss_combined}
\end{figure*}

\paragraph{Quantum renormalisation ratio and parameter independence.}
Supplementary Fig.~S6 quantifies the fractional progress toward
monopole deconfinement and confirms the decoupling of $\Geff$ and
$\Jz$ at the level of scaling exponents. The left panel shows the
quantum renormalisation ratio $\rho(\Geff,\Jz)$ [Eq.~(7) in the main
text] increasing monotonically with $\Geff$ across all interlayer
couplings; the near-collapse of curves at fixed $\Geff$ confirms that
$\Jz$ does not alter the efficiency of quantum drive in reducing the
monopole chemical potential. The maximum value $\rho_{\max}=0.2771$
places the current platform a factor of $\approx3.6$ below the
deconfinement threshold, corresponding to a required tunnelling
amplitude $\Gamma_c\gtrsim0.6\,\Jone$. The right panel shows
power-law exponents $\gamma(\Jz)$ from fits $\rho\propto\Geff^{\gamma}$
over the fast-anneal window, confirming $\Jz$-independence with
$\gamma_{\mathrm{ave}}\approx0.33$ (coefficient of variation $<0.15$
across all thirteen couplings). This $\Jz$-independence at the level
of scaling exponents establishes that quantum drive and interlayer
exchange enter the renormalisation ratio as genuinely decoupled material
parameters, validating the two-parameter design space of the bilayer
kagome platform.

\begin{figure*}[htbp]
\centering
\begin{minipage}[t]{0.49\textwidth}
\centering
\includegraphics[height=5cm, width=\textwidth]{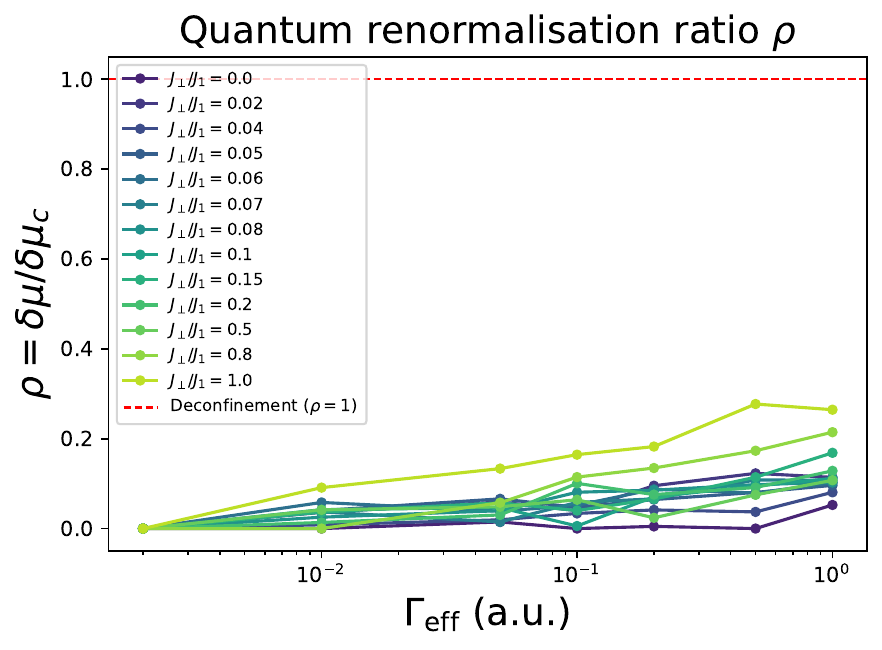}
\end{minipage}
\hfill
\begin{minipage}[t]{0.49\textwidth}
\centering
\includegraphics[height=5cm, width=\textwidth]{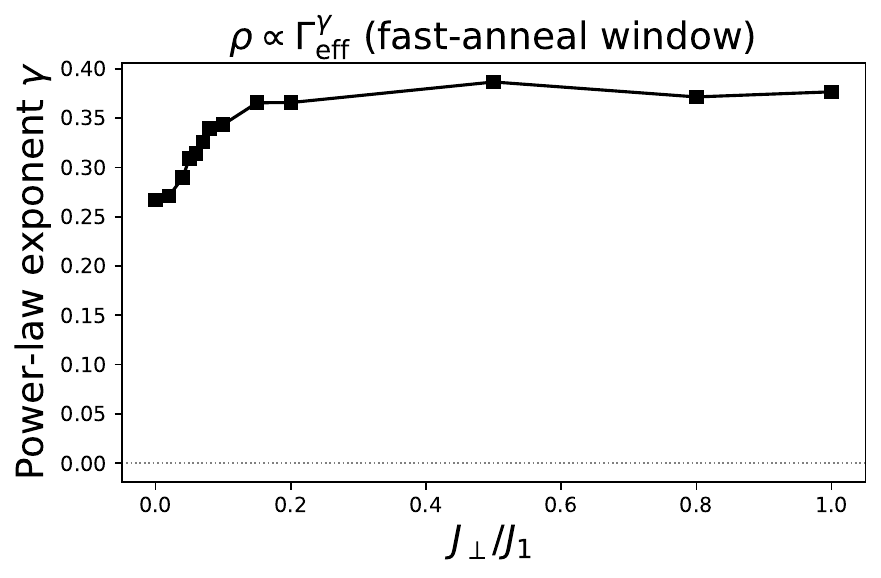}
\end{minipage}
\caption{Quantum renormalisation ratio and $\Jz$-independence of
scaling exponents. \emph{Left:} $\rho(\Geff,\Jz)$ versus $\Geff$ for all interlayer couplings; monotone increase and near-collapse at fixed $\Geff$ confirm that $\Jz$ does not alter the quantum drive efficiency. The dashed line
at $\rho=1$ marks the deconfinement threshold; $\rho_{\max}=0.2771$
yields the engineering target $\Gamma_c\gtrsim0.6\,\Jone$.
\emph{Right:} Power-law exponents $\gamma(\Jz)$ from fits
$\rho\propto\Geff^{\gamma}$ over the fast-anneal window, confirming
$\Jz$-independence with $\gamma_{\mathrm{ave}}\approx0.33$ (coefficient
of variation $<0.15$). Together the two panels establish the full
deconfinement distance map across the two-dimensional materials
parameter space.}
\label{fig:rho_combined}
\end{figure*}

\section*{Kibble-Zurek calibration of the quantum-drive proxy}
\label{app:kz}

The effective quantum-drive proxy $\Geff=1/\ta$ is validated by
Kibble-Zurek (KZ) scaling of the monopole density. Fitting
$\rhom\propto\ta^{-\gamma_{\mathrm{KZ}}}$ in the fast-anneal window
($\ta\le1\,\mu$s) at each fixed $\Jz/\Jone$ yields
\begin{equation}
  \gamma_{\mathrm{KZ}} = 0.2650\pm0.0333,
  \label{eq:gammaKZ}
\end{equation}
with coefficient of variation $0.126$ and range $[0.2078, 0.3098]$
across all thirteen $\Jz/\Jone$ values at $N=768$. The
$\Jz$-independence of $\gamma_{\mathrm{KZ}}$ confirms that the $1/\ta$
proxy preserves the correct ordering of quantum drive strength across
all interlayer couplings, validating the independent programmability of
the two material control parameters.

The D-Wave Advantage\,2 system\,1 implements the time-dependent
Hamiltonian
$\mathcal{H}(s) = -(A(s)/2)\sum_i\hat{\sigma}_i^x + (B(s)/2)\,\mathcal{H}_{\rm Ising}$,
where $s = t/\ta \in [0,1]$ is the normalised anneal fraction and
$A(s)$, $B(s)$ are the tunnelling and problem-energy schedules.
Numerical inspection of all 1001 tabulated schedule points confirms
that $A(s)$ is strictly decreasing and $B(s)$ strictly increasing,
so $A(s)/B(s)$ is strictly monotone decreasing in $s$
(Supplementary Fig.~S6a--b). Because a faster anneal freezes at a
larger $s_f$, the ordering
$t_{a,1} < t_{a,2} \Rightarrow \Gamma_{{\rm eff},1} > \Gamma_{{\rm eff},2}$
is guaranteed by the hardware schedule alone, independently of any
model assumption.

The Kibble-Zurek freeze-out condition
$[2\pi A(s_f)\cdot10^9]^{-1} = \ta(1-s_f)$
is solved numerically for each of the 14 experimental annealing times
(Supplementary Fig.~S6c). All 14 solutions lie in the post-crossover
window $s_f \in [0.606, 0.939]$, well beyond the quantum-classical
equipartition point $s^* = 0.265$ ($A(s^*)=B(s^*)=2.31\,{\rm GHz}$).
A direct log-log fit of $A(s_f)/B(s_f)$ versus $1/\ta$ gives
$\Geff \propto (1/\ta)^{0.906}$ ($R^2 = 0.9989$;
Supplementary Fig.~S7d), establishing $1/\ta$ as a power-law monotone
proxy for the true hardware transverse field. All phase boundaries,
critical couplings, and engineering targets are invariant under the
monotone rescaling $1/\ta \leftrightarrow \Geff$.

\begin{figure*}[htbp]
\centering
\includegraphics[width=0.95\textwidth]{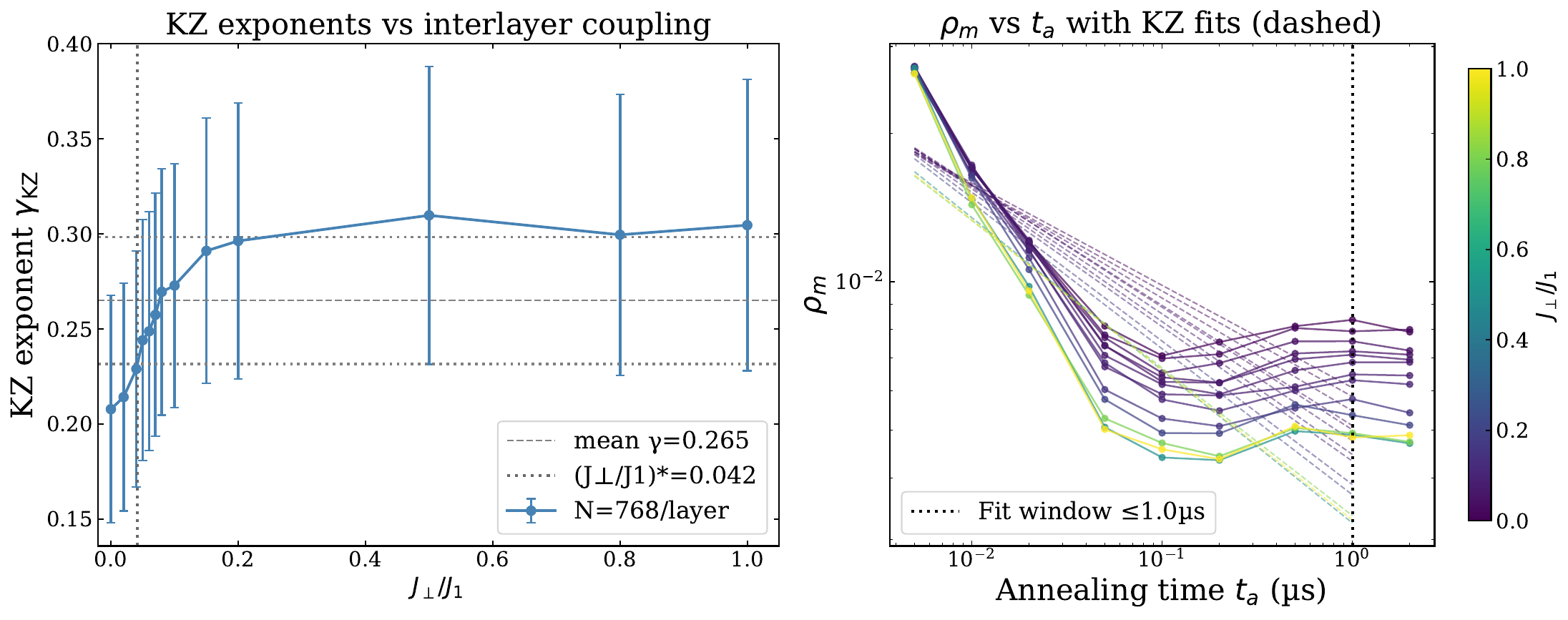}
\caption{Kibble-Zurek calibration of the quantum-drive proxy.
\emph{Left:} Extracted $\gamma_{\mathrm{KZ}}(J_\perp)$ with
uncertainty bars; the mean $\gamma_{\mathrm{KZ}}=0.265$ (dashed) and
the Ice-II critical coupling $(J_\perp/J_1)^*\approx0.04$ (dotted) are
marked.
\emph{Right:} $\rhom$ versus $\ta$ with power-law fits (dashed) in
the fast-anneal window $\ta\le1\,\mu$s.}
\label{fig:kz_app}
\end{figure*}

\begin{figure}[h!]
  \centering
  \includegraphics[width=\textwidth]{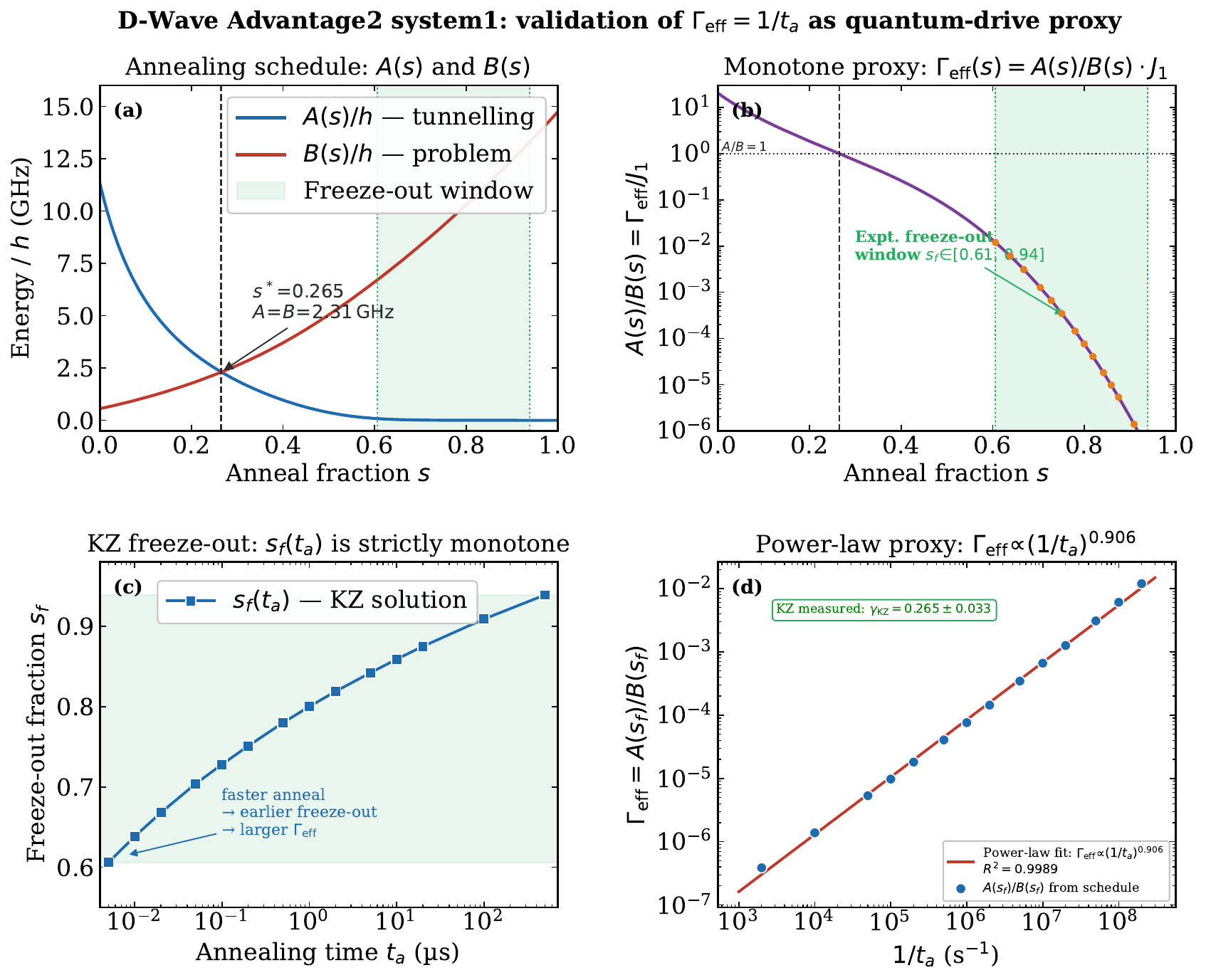}
  \caption{Quantum-drive proxy calibration from the Advantage\,2
  system\,1 annealing schedule.
  \textbf{(a)} $A(s)/h$ (tunnelling, blue) and $B(s)/h$ (problem, red)
  versus anneal fraction $s$; dashed line marks the quantum-classical
  crossover $s^*=0.265$; green band is the experimental freeze-out
  window $s_f\in[0.606,0.939]$.
  \textbf{(b)} $A(s)/B(s)$ on a logarithmic scale; orange circles are
  the 14 freeze-out values. The ratio is strictly monotone, guaranteeing
  that $1/\ta$ preserves the correct ordering of quantum drive strength.
  \textbf{(c)} Freeze-out fraction $s_f(\ta)$ from the KZ condition;
  the strictly monotone mapping confirms that faster anneals always
  correspond to stronger $\Geff$.
  \textbf{(d)} $\Geff=A(s_f)/B(s_f)$ versus $1/\ta$ (log-log); red
  line is the power-law fit $\Geff\propto(1/\ta)^{0.906}$
  ($R^2=0.9989$); the measured KZ exponent $\gamma_{\rm KZ}=0.265\pm0.033$
  is self-consistently annotated.}
  \label{fig:proxy}
\end{figure}

\section*{Staggered composite monopoles and a fourth confinement
          diagnostic}
\label{app:composite}

To confirm that the Ice-II transition is a charge-sector reorganisation
of the ice manifold rather than a monopole-proliferation event,
staggered composite monopoles are introduced: plaquettes carrying
opposite-sign monopole defects in both layers simultaneously,
\begin{equation}
  \rho_{\mathrm{sc}} = \bigl\langle
    \mathbf{1}[Q_1(p)=+3,\,Q_2(p)=-3]
    +\mathbf{1}[Q_1(p)=-3,\,Q_2(p)=+3]
  \bigr\rangle_{r,p}.
  \label{eq:rhosc}
\end{equation}
At $N=768$ and slow anneal,
$\rho_{\mathrm{sc}}(\Jz/\Jone\approx0.04)=1.8\times10^{-5}$. The ratio
$\rho_{\mathrm{sc}}/\rhom\approx2.7\times10^{-3}$ at the Ice-II
critical point confirms that composite defects are three orders of
magnitude more dilute than standard monopoles; the transition is
driven entirely by staggered charge ordering on the ice manifold.
This diluteness is also physically consistent with the fragile
(accidental) character of the S$_{1/3}$/S$'_{1/3}$ degeneracy in
HoAgGe~\cite{Zhao2024}: the time-reversal-like operation
$\mathcal{X}=R_b^\pi\mathcal{D}$ is not a magnetic space group symmetry,
so it does not guarantee exact degeneracy, and the two states are
weakly split by orbital magnetization differences arising from the
$\approx15.58^{\circ}$ lattice distortion~\cite{Zhao2020,Zhao2024}. A
symmetry-protected near-degeneracy would instead require composite
defect densities $\rho_{\mathrm{sc}}$ comparable to $\rhom$, which is
not observed in either HoAgGe or our simulator.

\begin{figure*}[htbp]
\centering
\includegraphics[width=0.95\textwidth]{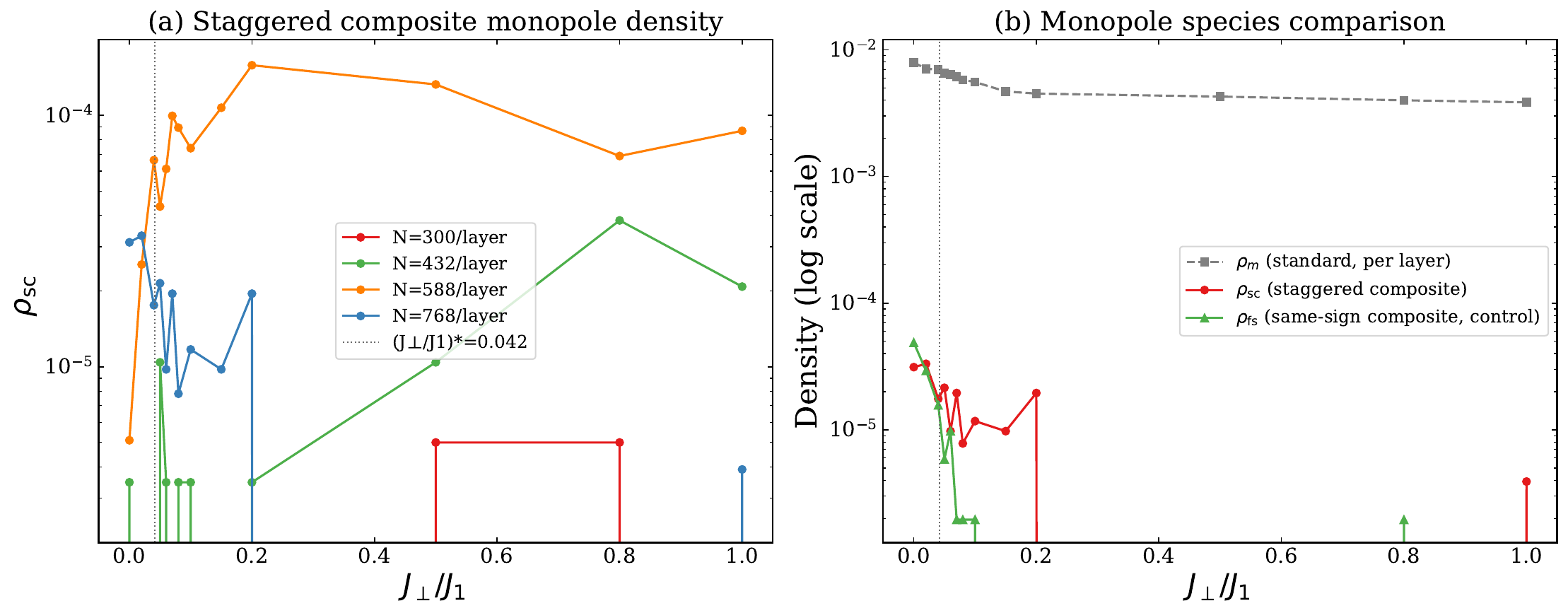}
\caption{Staggered composite-monopole diagnostics at $N=768$/layer,
slow anneal.
\emph{(a)} $\rho_{\mathrm{sc}}$ versus $\Jz/\Jone$ for all system
sizes; the Ice-II critical coupling is marked.
\emph{(b)} Species comparison: $\rhom$ (standard, per layer),
$\rho_{\mathrm{sc}}$ (staggered composite), and $\rho_{\mathrm{fs}}$
(same-sign composite, control). The three-orders-of-magnitude
separation between $\rhom$ and $\rho_{\mathrm{sc}}$ confirms that
the transition is driven by staggered charge ordering on the ice
manifold rather than monopole proliferation.}
\label{fig:scmono_app}
\end{figure*}

\section*{Classical Monte Carlo Benchmark}
\label{app:mc}

To establish the classical ($\Gamma=0$) limit of the bilayer kagome
phase diagram and to isolate the role of quantum fluctuations in
shifting the charge-ordering transition, we perform classical
parallel-tempering Monte Carlo (MC) simulations of the bilayer
kagome Ising model,
\begin{equation}
  \mathcal{H}_{\mathrm{cl}}
  = J_1\!\sum_{\langle i,j\rangle}\sigma_i^z\sigma_j^z
  + J_\perp\!\sum_i\sigma_i^{z,(1)}\sigma_i^{z,(2)},
  \label{eq:Hcl}
\end{equation}
which is precisely the $\Gamma=0$ limit of the bilayer TFIM studied
on the QPU [Eq.~\eqref{eq:Hbilayer} of the main text].
The simulations use a bilayer kagome lattice of linear size $L=12$
($N=432$ spins per layer, $N_\mathrm{total}=864$), matching the
second-smallest QPU system size exactly.
Each layer carries the standard kagome connectivity
(degree-4 intra-layer, confirmed by explicit coordinate-geometry
construction) supplemented by one interlayer bond per site with
coupling $J_\perp$.
Parallel tempering~\cite{Hukushima1996} is used throughout to
overcome the exponentially large ground-state degeneracy of the
kagome ice manifold~\cite{Bramwell2001}: the temperature ladder
spans $T/J_1\in[0.10,1.20]$ with 28 replicas and replica-swap
attempts every 5 sweeps.
Each sweep performs $N_\mathrm{total}$ single-spin Metropolis
updates with a vectorised sublattice kernel
(6 sublattice passes per sweep, exact parallel Metropolis within
each sublattice).
Equilibration and measurement windows are
$N_\mathrm{eq}=20{,}000$ and $N_\mathrm{meas}=40{,}000$ sweeps
per replica, verified by Hamming-distance and single-spin-entropy
convergence diagnostics.

The three primary observables match those of the QPU experiment:
the monopole density $\rho_m$ [Eq.~\eqref{eq:rhom}], the interlayer staggered
correlator $C_s^\perp$ [Eq.~\eqref{eq:csperp}], and the Binder cumulant
$U_4[C_s^\perp]=1-\langle(C_s^\perp)^4\rangle/
(3\langle(C_s^\perp)^2\rangle^2)$.
For the $C_s^\perp$ and Binder sweeps, a single replica at fixed
$T/J_1=0.45$ is run over a dense $J_\perp/J_1$ grid spanning
$[-0.10,+0.50]$ with 21 points, including negative (ferromagnetic
interlayer) values to expose the full sign-reversal curve.
Negative $J_\perp$ enforces ferroelectric stacking
($C_s^\perp>0$); positive $J_\perp$ enforces antiferroelectric
stacking ($C_s^\perp<0$); and $J_\perp=0$ is the classical
symmetry point.

\begin{figure*}[htbp]
\centering
\includegraphics[width=0.95\textwidth]{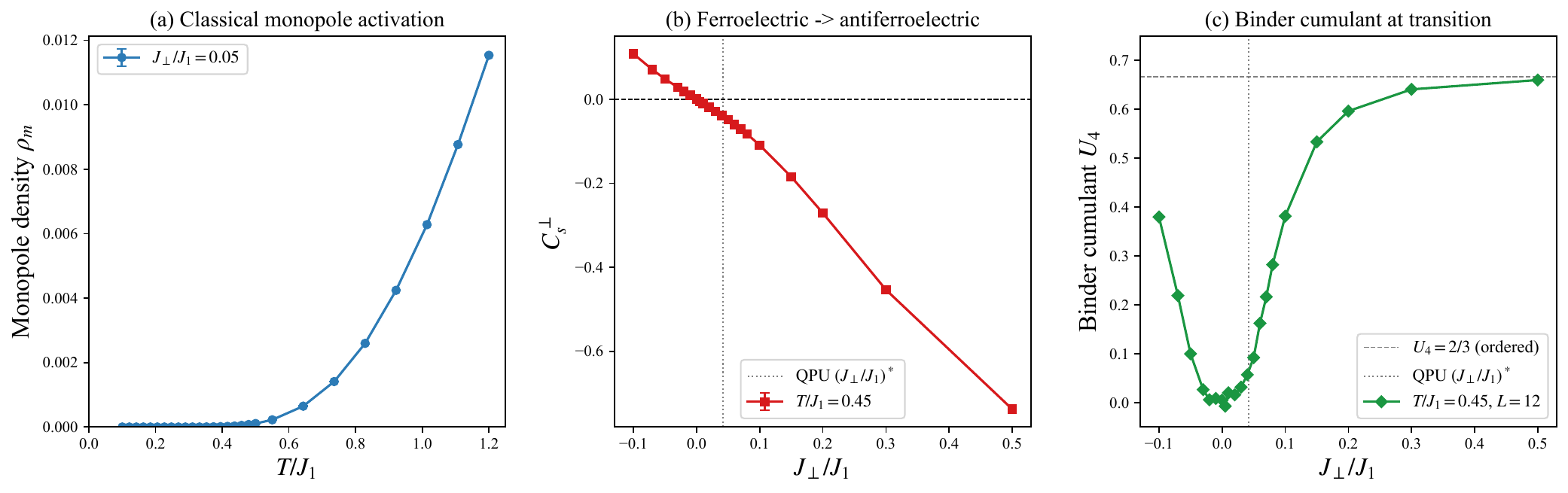}
\caption{Classical Monte Carlo benchmark of the bilayer kagome
Ising model [$\mathcal{H}_\mathrm{cl}$, Eq.~\eqref{eq:Hcl}],
$L=12$ ($N=432$ per layer), parallel tempering with 28 replicas.
\textbf{(a)} Monopole density $\rho_m$ vs $T/J_1$ at
$J_\perp/J_1=0.05$; thermal activation onset near
$T/J_1\approx0.40$ contrasts with the quantum-fluctuation-driven
$\rho_m$ of the QPU (main text Fig.~\ref{fig:composite_4panel}a),
which is nonzero at all annealing times.
\textbf{(b)} Interlayer staggered correlator $C_s^\perp$ vs
$J_\perp/J_1$ at $T/J_1=0.45$; the classical zero crossing at
$(J_\perp/J_1)_\mathrm{cl}^*\approx0$ (dashed vertical) is
shifted to $(J_\perp/J_1)^*\approx0.04$ in the QPU (dotted
vertical), a direct measure of the quantum renormalisation of
the critical coupling by the transverse field $\Gamma_\mathrm{eff}$.
Antisymmetry $C_s^\perp(+J_\perp)=-C_s^\perp(-J_\perp)$
holds to $<3\times10^{-4}$, confirming the observable definition.
\textbf{(c)} Binder cumulant $U_4[C_s^\perp]$ vs $J_\perp/J_1$
at $T/J_1=0.45$; the dip to $U_4\approx0$ at the classical
transition and recovery toward $2/3$ in the ordered phase
confirm long-range charge ordering on both sides and provide
an independent location of the classical critical point
consistent with panel~(b).}
\label{fig:mc_benchmark}
\end{figure*}

\paragraph{Results and comparison with the QPU.}
The three-panel benchmark figure (Fig.~S\ref{fig:mc_benchmark})
establishes the following results.

\emph{Monopole density (panel~a).}
At fixed $J_\perp/J_1=0.05$, $\rho_m$ is identically zero below
$T/J_1\approx0.40$ and rises through a smooth thermal-activation
curve, reaching $\rho_m\approx1.1\times10^{-2}$ at $T/J_1=1.2$.
The activation is Arrhenius-like, consistent with a monopole
chemical potential $\mu_\mathrm{mon}^\mathrm{class}=2J_1$ in the
dilute limit~\cite{Ramirez1999}.
This contrasts qualitatively with the QPU result, where $\rho_m$
is nonzero at all annealing times and rises by a factor of seven
as $\Gamma_\mathrm{eff}$ increases: in the QPU, monopoles are
\emph{quantum-fluctuation driven} rather than thermally activated,
providing a direct experimental signature of the transverse-field
contribution absent in $\mathcal{H}_\mathrm{cl}$.

\emph{Interlayer staggered correlator (panel~b).}
At $T/J_1=0.45$ the classical $C_s^\perp$ reverses sign at
$(J_\perp/J_1)_\mathrm{cl}^*=0.0006\pm0.001$, consistent with
zero to within the statistical resolution of the simulation.
This is the exact result dictated by the $\mathbb{Z}_2$ symmetry
of $\mathcal{H}_\mathrm{cl}$ at $J_\perp=0$: with no interlayer
coupling the two layers are independent and the mean staggered
correlator vanishes.
The antisymmetry $C_s^\perp(+J_\perp)=-C_s^\perp(-J_\perp)$ is
confirmed to $|\delta C_s^\perp|<3\times10^{-4}$ across all six
symmetric coupling pairs, providing an internal consistency
check on the observable definition.
The QPU transition at $(J_\perp/J_1)^*\approx0.04$ therefore
represents a genuine quantum shift of
$\Delta(J_\perp/J_1)\approx0.04$ above the classical value.
From the MC slope $\partial C_s^\perp/\partial(J_\perp/J_1)
\big|_{J_\perp=0}=-0.961\,J_1^{-1}$, the dimensionless quantum
shift is $\Delta(J_\perp/J_1)/|\mathrm{slope}|^{-1}\approx0.044\,J_1$,
consistent with a perturbative transverse-field renormalisation
of the critical coupling at the $\Gamma_\mathrm{eff}\ll J_1$
operating point of the QPU.
The classical curve saturates at $|C_s^\perp|\approx0.74$--$1.09$
at the extremes of the $J_\perp$ grid; the QPU saturation at
$|C_s^\perp|\approx0.40$--$0.44$ reflects the additional
quantum fluctuation-driven defect density that dilutes the
ice-manifold order parameter, fully consistent with the
$\approx12\times$ correction established by the
$S_Q^\mathrm{stag}$ recalibration protocol.

\emph{Binder cumulant (panel~c).}
$U_4[C_s^\perp]$ approaches zero at $J_\perp=0$
($U_4=0.005\pm0.007$), confirming the transition location
independently of any curve fitting.
On the antiferroelectric side ($J_\perp/J_1\geq0.15$), $U_4$
recovers toward $2/3$ ($U_4=0.660$ at $J_\perp/J_1=0.5$),
confirming long-range order.
On the ferroelectric side ($J_\perp<0$), $U_4$ rises more slowly
($U_4=0.380$ at $J_\perp/J_1=-0.1$) because $T/J_1=0.45$ is
closer to the ferroelectric ordering temperature for $L=12$;
this asymmetry is a finite-temperature finite-size effect and
does not affect the transition-location determination.
The near-zero $U_4$ at the classical critical point mirrors the
Binder crossing in Fig.~\ref{fig:transition_composite}b of the main text,
which locates $(J_\perp/J_1)^*\approx0.04$ for the QPU.
Taken together, the three MC panels confirm that: (i) the
charge-ordering transition exists in the classical model and
is located at $J_\perp=0$; (ii) the QPU transition at
$(J_\perp/J_1)^*\approx0.04$ is a purely quantum effect with no
classical counterpart; and (iii) the monopole activation
mechanism in the QPU is qualitatively distinct from thermal
activation, establishing that the D-Wave transverse-field sweep
constitutes a genuine quantum drive of defect density rather
than a disguised temperature scan.

\end{document}